\documentclass[useAMS,usenatbib]{mn2e}
\usepackage[T1]{fontenc}
\usepackage{aecompl}
\usepackage{upgreek}

\usepackage[draft]{hyperref}

\usepackage{graphicx}
\usepackage{multirow}
\usepackage{amssymb}
\usepackage{hyperref}
\hypersetup{colorlinks=false,
		linkcolor=black,
		urlcolor=cyan}

\title[SN 2012ca]{On type IIn/Ia-CSM supernovae as exemplified by SN 2012ca\thanks{Based on observations collected at the European Organisation for Astronomical Research in the Southern Hemisphere, Chile, as part of programme 188.D-3003 (PESSTO).} }
\author[C. Inserra et al.]{C. Inserra$^{1}$\thanks{E-mail: c.inserra@qub.ac.uk(CI)},
M. Fraser$^{2}$, S. J. Smartt$^{1}$,
S. Benetti$^{3}$, T.-W. Chen$^{4}$, M. Childress$^{5}$,
\newauthor A. Gal-Yam$^{6}$, D. A. Howell$^{7,8}$, T. Kangas$^{9}$,G. Pignata$^{10,11}$,  J. Polshaw$^{1}$, M. Sullivan$^{12}$,
\newauthor  K. W. Smith$^{1}$, S. Valenti$^{7,8}$, D. R. Young$^{1}$, S. Parker$^{13}$ T. Seccull$^{1}$ and M. McCrum$^{1}$
\\
$^{1}$Astrophysics Research Centre, School of Mathematics and Physics, Queen's University
  Belfast, Belfast BT7 1NN, UK\\
$^{2}$Institute of Astronomy, University of Cambridge, Madingley Rd., Cambridge, UK\\
$^{3}$INAF - Osservatorio Astronomico di Padova, Vicolo dell'Osservatorio 5, I-35122 Padova, Italy\\
$^{4}$Argelander Institute for Astronomy, University of Bonn, Auf dem H\"ugel 71, D-53121 Bonn, Germany\\
$^{5}$Research School of Astronomy and Astrophysics, The Australian National University, Weston Creek, ACT 2611, Australia\\
$^{6}$Benoziyo Center for Astrophysics, Weizmann Institute of Science, 76100 Rehovot, Israel\\
$^{7}$Las Cumbres Observatory Global Telescope Network, 6740 Cortona Dr, Suite 102, Goleta, CA 93117, USA\\
$^{8}$Department of Physics, University of California, Santa Barbara, Broida Hall, Mail Code 9530, Santa Barbara, CA 93106-9530, USA\\
$^{9}$Tuorla Observatory, Department of Physics and Astronomy, University of Turku, V\"ais\"al\"antie 20, FI-21500 Piikki\"o, Finland\\
$^{10}$Departamento de Ciencias Fisicas, Universidad Andres Bello, Avda. Republica 252, Santiago, Chile\\
$^{11}$Millennium Institute of Astrophysics, Chile\\
$^{12}$School of Physics and Astronomy, University of Southampton, Southampton, SO17 1BJ, UK\\
$^{13}$Backyard Observatory Supernova Search, Oxford, Canterbury, New Zealand}

\def\kms{km\,s$^{-1}$}
\def\Ha{H{$\alpha$}}

\def\ni{$^{56}$Ni}
\def\co{$^{56}$Co}
\def\fe{$^{56}$Fe}

\def\M{M$_{\odot}$}

\def\ca{SN~2012ca}

\begin{document}

\date{Received.....; accepted...........}

\pagerange{\pageref{firstpage}--\pageref{lastpage}} \pubyear{}

\maketitle

\label{firstpage}

\begin{abstract}

We present the complete set of ultra-violet, optical and near-infrared photometry and spectroscopy for SN 2012ca, covering the period from 6~days prior to maximum light, until 531 days after maximum. The spectroscopic time series for SN 2012ca is essentially unchanged over 1.5~years, and appear to be dominated at all epochs by signatures of interaction with a dense circumstellar medium rather than the underlying supernova (SN). At late phases, we see a near infrared excess in flux which is possibly associated with dust formation, although without any signs of accompanying line shifts. SN 2012ca is a member of the set of type of the ambiguous IIn/Ia-CSM SNe, the nature of which have been debated extensively in the literature. The two leading scenarios are either a type Ia SN exploding within a dense CSM from a non-degenerate, evolved companion, or a core-collapse SN from a massive star. While some members of the population have been unequivocally associated with type Ia SNe, in other cases the association is less certain. While it is possible that \ca\/ does arise from a thermonuclear SN, this would require a relatively high (between 20 and 70 per cent) efficiency in converting kinetic energy to optical luminosity, and a massive ($\sim2.3-2.6$ \M) circumstellar medium. On the basis of energetics, and the results of simple modelling, we suggest that \ca\/ is more likely associated with a core-collapse SN. This would imply that the observedset of similar SNe to SN~2012ca is in fact originated by two populations, and while these are drawn from physically distinct channels, they can have observationally similar properties.

 \end{abstract}
 

\section{Introduction}\label{sec:intro}

Supernovae (SNe) are the terminal catastrophic explosions of stars, and are produced by two main physical mechanisms.
The first is a thermonuclear explosion (SN Ia) when a CO white dwarf (WD) approaches  the
Chandrasekhar limit after accreting hydrogen and helium from a companion
star (the single degenerate channel), or through the
merger of two WDs \citep[the double degenerate channel;][]{hn00}. The second mechanism is the gravitational collapse of the core of a massive star (CC-SNe), which
will leave a compact remnant \citep{2012ARNPS..62..407J}. CC-SNe are classified according to the presence (SNe II) or absence (SNe I) of H in
their spectra; SNe I are further sub-divided into SNe Ib and SNe Ic depending on whether or not they show signs of He in their spectra \citep{fi97}. A relatively heterogeneous subclass of CC-SNe are termed type IIn, these show Balmer emission lines with composite profiles, including narrow emission components \citep{sc90} which are formed as the SN ejecta collides with a dense, H-rich circumstellar medium (CSM), and are sometimes referred to as ``interacting'' SNe.

In the last two decades several examples of peculiar interacting SNe have been discovered, which were originally associated with the class of SNe IIn. The first to be discovered was SN~1997cy \citep{ge00,tu00}, which showed multicomponent hydrogen lines superimposed on a continuum with broad SN features. Subsequently, SN~2002ic, which appeared quite similar to SN~1997cy, was shown to have pre-peak spectra that are not {\it dominated} by interaction\footnote{While SN 2002ic showed a narrow H$\alpha$ emission line in its earliest spectra, it was not until $\sim$2 weeks post maximum that the interaction contributed more flux than the underlying SN \protect\citep{ha03}.} and resembling those of bright type Ia such as SNe 1991T and 1999aa \citep{ha03,de04,wv04}, although it did not show the typical $I$-band secondary maximum seen in the lightcurves of type Ia SNe. Nevertheless, SN 2002ic 
was suggested to have a thermonuclear origin
and led to the introduction of the label ``Ia-CSM'' in order to identify 
these transients. Approximately a dozen similar SNe have been found to date which have similar absolute magnitudes and spectra to SNe 1997cy and 2002ic \citep{si13a}, although only one of these has shown spectra not dominated by interaction \citep[PTF11kx,][]{di12,si13b}. 

These SNe exhibits obvious signs of their ejecta
interacting with circumstellar material (CSM). Their spectra appear to be a ``diluted'' spectrum
of a bright SN Ia, along with superimposed  H
emission lines. PTF11kx shows the strongest evidence
for being a thermonuclear event interacting with CSM expelled by a
companion red giant star \citep{di12}, including the typical SN Ia $I$-band double peak.
The growing arguments that many such events with narrow hydrogen - that according to standard SN taxonomy should be classified as type IIn SNe - could be thermonuclear has resulted in SN~2008J \citep{ta12} and several other transients such as SN~2005gj \citep{al06,pr07} being labelled as Ia-CSM SNe. In most cases, this has been based on the similarity of their interaction-dominated spectra, as for example in the sample presented by \citet{si13a}. Following from this, \citet{le15} investigated how the flux ratio between the underlying SN and the continuum affected the spectroscopic classification of these events, and argued that the fraction of Type Ib/c SNe with interaction that are misclassified as SN 1991T-like is probably small.

However, not all authors have interpreted such SNe as necessarily being thermonuclear events, with \citet{be06} and \citet{tr08} 
arguing for a core-collapse origin for SNe 2002ic and 2005gj, respectively. 
These ambiguous and interacting events are important for
determining the possible progenitor channels for SNe Ia in particular,
but are rare. 

\ca\/ is the closest among these rare objects and the spectroscopic analysis - revealing O, C and He lines, together with a $\sim$200 \kms\/ \Ha\/ - led \citet{in14} to propose a core-collapse origin for the SN. \citeauthor{in14} found possible blueshifts in some of the identified lines, which we claimed could be accounted for by an asymmetric explosion and ejecta. On the other hand, \citet{fox15} disputed the identification of some of the intermediate mass element lines, and suggested a thermonuclear origin for \ca. In particular, \citeauthor{fox15} noted that the low [Fe~{\sc iii}]/[Fe~{\sc ii}] ratio and strong [Ca~{\sc ii}] were similar to what is observed in super-Chandrasekhar mass candidates such as SN~2009dc \citep{tau11}. \citeauthor{fox15} also identified high-ionization coronal lines in the late time spectra of \ca, which they suggest arise from the CSM external to the shock front, and were excited by x-rays from ongoing interaction.

Since there is unambiguous evidence for a thermonuclear origin for only a single object (namely PTF11kx) we will refer hereafter to \ca\/ and objects showing similar observables as type IIn/Ia-CSM. Here we report the full dataset collected by the Public ESO Spectroscopy Survey of Transient Objects (PESSTO)\footnote{www.pessto.org} together with additional data (Sections~\ref{sec:obs},~\ref{sec:ph},~\ref{sec:bol} and~\ref{sec:sp}), making \ca\/ the best sampled object of this peculiar group of SNe. Section~\ref{sec:dis} is devoted to the analysis of the spectrophotometric data,  in order to understand the origin of \ca\/ and similar objects, and to investigate if they arise from a single progenitor channel or whether distinct sub-groups can be identified. Finally, a short summary is reported in Section~\ref{sec:end}.

\begin{figure}
\includegraphics[width=9.2cm]{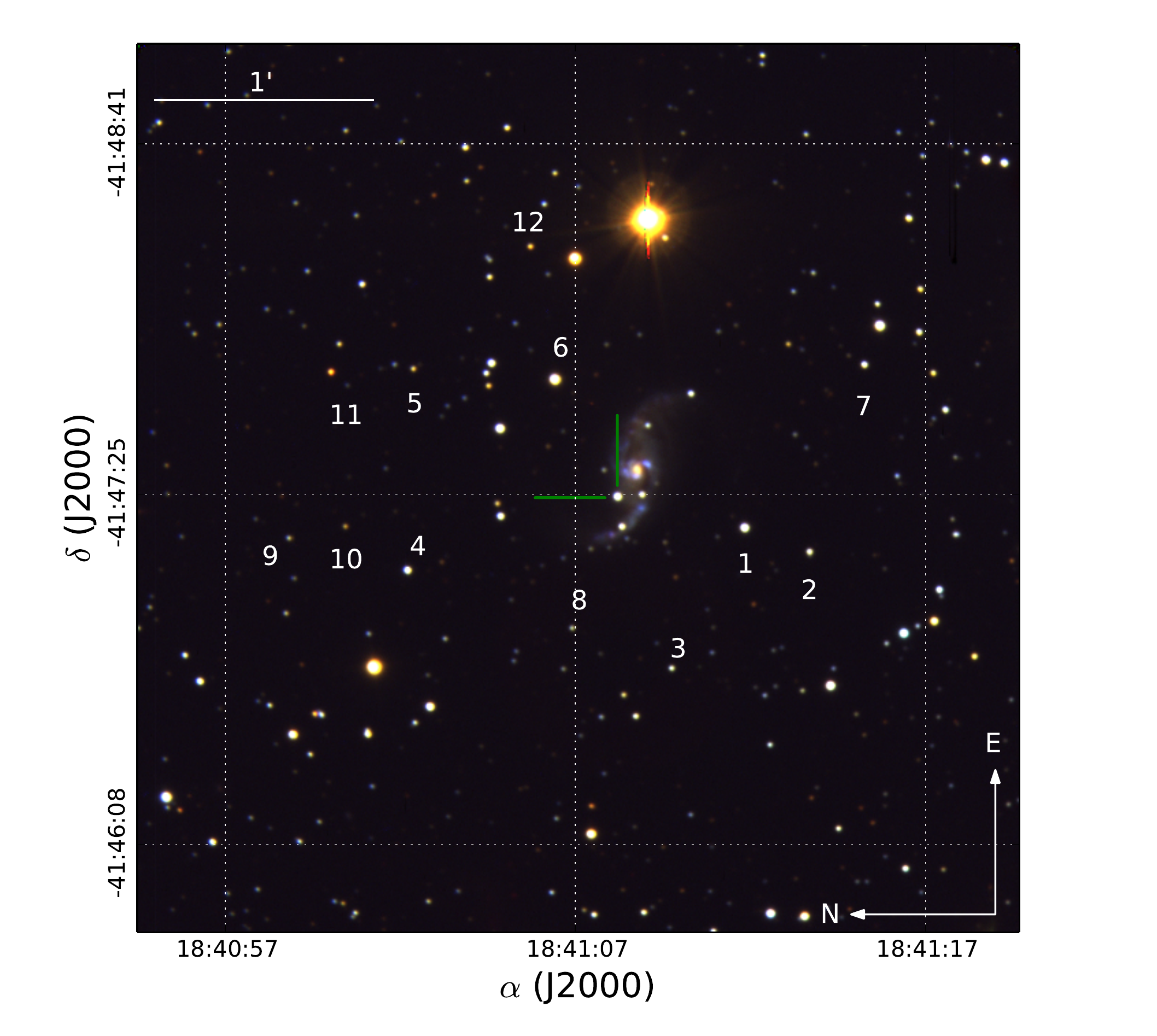}
\caption{NTT+EFOSC2 {\it Ugr} filter colour composite of \ca\/ in ESO 336-G009. The SN position is indicated with green cross marks. The sequence of stars in the field used to calibrate the optical and NIR magnitudes of \ca\/ is indicated.}
\label{fig:fc}
\end{figure}

\section{Observations and data reduction}\label{sec:obs}

\ca\/ (PSN J18410706-4147374) was discovered in the late-type spiral (SABc) galaxy ESO 336-G009 by Parker on {\sc ut} 2012 April 25.6 at an unfiltered magnitude of m $\sim 14.8$ mag \citep{dr12}.
We performed our own photometry on the discovery image and find it to be at {\it R}=14.94$\pm$0.03 mag, as reported in Tab.~\ref{table:snm} (see Sect. \ref{sec:ph} for details of the photometry). The object coordinates have been measured on our astrometrically
calibrated images to be $\alpha = 18^{\rm h}41^{\rm m}07^{\rm s}.21 \pm0^{\rm s}.05$, $\delta = -41^{\rm o}47'03''.01 \pm0''.05$ (J2000).
\ca\/ is located in an inner region outside the spiral arms of the host, 4.5 arcsec N/W of the centre of the host galaxy (Fig.~\ref{fig:fc}),
which corresponds to a projected distance of $\sim 1.7$ kpc from the nucleus for an adopted distance to ESO 336-G009
of $\sim80$ Mpc (see Section~\ref{ss:dr}).
A spectrum taken by PESSTO at the New Technology
Telescope (NTT) + ESO Faint Object Spectrograph and Camera 2 (EFOSC2) \citep{va12,in12} on April 29.4 {\sc ut}  showed that \ca\/ was a Type IIn SN resembling SN~1997cy $\sim$60 d after maximum.

\subsection{Host galaxy properties}\label{ss:dr}

NED\footnote{NASA/IPAC Extragalactic Database} lists a heliocentric radial velocity of $v_{\rm hel} = 5834 \pm37$ \kms\ for ESO~336-G009.
This is consistent with the redshift $z=0.019$ as measured from the narrow Balmer emission lines associated with SN 2012ca.
Adopting a standard cosmology with
${\rm H}_{0}=72$ \kms\/, ${\rm \Omega_{M}}=0.27$ and ${\rm \Omega_{\lambda}}=0.73$, the distance modulus for \ca\ is 34.52~mag (80.3 Mpc). 

The foreground Galactic reddening toward ESO 336-G009 is $E_{\rm g}(B-V) = 0.06$ mag from the \cite{sf11} dust maps.
The available SN spectra do not show Na~{\sc id} lines from the host galaxy hence we adopt
a total reddening along the line of sight towards \ca\/ of $E_{\rm tot}(B-V) = 0.06$ mag.

In our last spectrum taken with NTT+EFOSC2 on 2013 October 13 no flux from the continuum or pseudo-continuum of \ca\ was detected (see Section~\ref{sec:sp}). However the \Ha\/ features appears to be broader than expected for an unresolved line, possibly due in part to blending with [N~{\sc ii}] $\lambda$6584. However, after deblending the lines, the \Ha\/ component still appears too broad (v$\sim$1000 \kms\/), and is likely still contaminated by the SN/SN+interaction contribution. While we were hence unable to measure the \Ha\/ flux of the host galaxy at the position of the SN, if we take the measured flux as an upper limit we can still derive an upper limit to the star formation rate (SFR) of the host galaxy at the SN location. We measured a flux F$_\mathrm{Gal}$(\Ha)$\lesssim7.30\times10^{-15}$ (EW(\Ha)~$\lesssim$~45.3~\AA\/) which, given the distance of 80.3 Mpc, corresponds to $L({\sc H}\alpha)\lesssim8.65\times10^{40}$ erg s$^{-1}$ and a star formation rate (SFR)~$\lesssim$~0.7 \M\/ year$^{-1}$, assuming solar metallicity and applying the calibration of \citet{ken94} and \citet{mad98}.

Type IIn/Ia-CSM have been observed both in late-type, mostly spiral, galaxies like \ca\/, SN~2008J and PTF11kx or in low-luminosity, low-metallicity environments such as SNe 1997cy, 2002ic, 2005gj. To date, no Type IIn/Ia-CSM have been found in early-type galaxies. These environments point towards a relatively young stellar population similar to those observed for core-collapse SNe, along with SN~1991T-like SNe~Ia \citep{si13a}.

\begin{figure*}
\includegraphics[width=18cm]{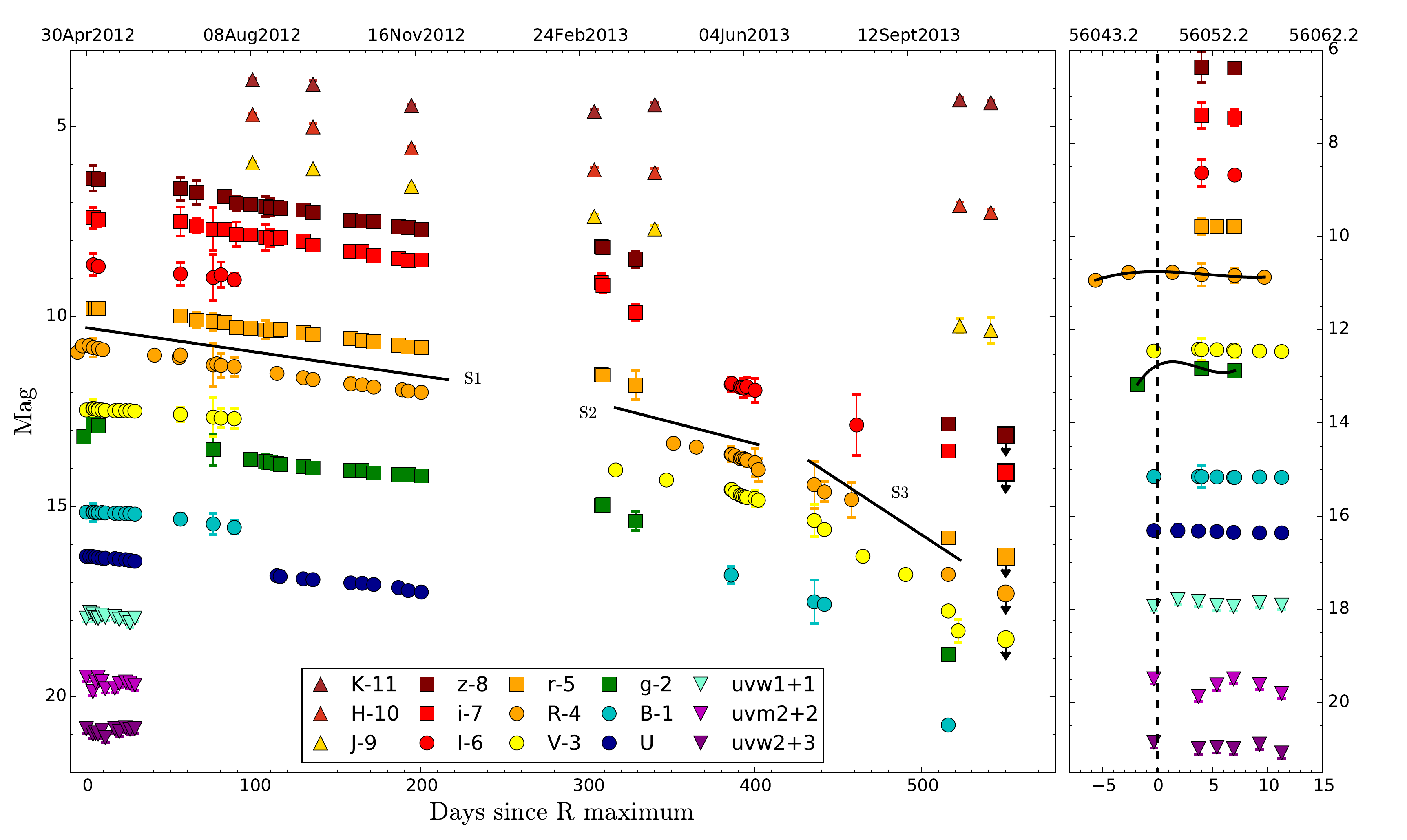}
\caption{Synoptic view of the light curves of \ca\/ in all available bands. The shifts from the original values reported in Tables~\ref{table:snm}~\&~\ref{table:sns} are indicated in the legend. The three distinct phases of decline seen in the lightcurve, and discussed in Sect. \ref{sec:ph} are labelled as S1, S2 and S3. The right panel shows a magnified region around the {\it R}-band maximum, which is indicated with a dashed line.}
\label{fig:lc}
\end{figure*}

\subsection{Data}
Optical spectro-photometric follow-up was mostly obtained with the NTT+EFOSC2 as part of PESSTO. The EFOSC2 photometry was obtained using {\it UBVR+griz} filters, and was taken on the same nights as EFOSC2 spectroscopy. The EFOSC2 data were augmented by observations from other ground-based telescopes including the Panchromatic Robotic Optical Monitoring and Polarimetry Telescopes \citep[PROMPT;][]{re05} PROMPT3 and PROMPT5 + {\it BVRIgriz} and the Small \& Moderate Aperture Research Telescope System (SMARTS)\footnote{Operated by the SMARTS Consortium} + ANDICAM + {\it RI}, which are both at Cerro Tololo; the 2m Faulkes South Telescope (FST) + MEROPE + {\it BVRI} at Siding Spring and the 1m telescopes + SINISTRO + {\it BVRI} at Cerro Tololo and Sutherland of the Las Cumbres Observatory Global Telescope Network \citep[LCOGT,][]{br13}. Supplementary unfiltered data were provided by Stu Parker of the Backyard Observatory Supernova Search (BOSS) using a 0.35 m Celestron equipped with a SBIG ST-10XME camera, located at Parkdale Observatory, Canterbury. Additional optical and ultraviolet (UV) photometry was obtained with the {\it Swift} satellite + UltraViolet and Optical Telescope (UVOT) (programme ID 32453; P.I. Inserra). The EFOSC2 images were reduced (trimmed, bias subtracted and flat-fielded) using the PESSTO pipelines \citep{sm14} while the PROMPT, SMARTS and LCOGT images were reduced automatically by their respective instrument-specific pipelines.

Photometric zero-points and colour terms were computed using observations of Landolt standard fields \citep{la92}, 20 of the 52 nights on which \ca\ were observed were photometric. Using the photometry on these nights we calibrated the magnitudes of a local stellar sequence, shown in Fig.~\ref{fig:fc}. The magnitudes of the local sequence stars are reported in Tables~\ref{table:ssj}~\&~\ref{table:sss} along with their r.m.s. (in brackets). Finally, the average magnitudes of the local-sequence stars were used to calibrate the photometric zero-points obtained on non-photometric nights or where the colour terms were not determined from Landolt standard fields.
 
The near-infrared (NIR) images were obtained with NTT + SOn oF Isaac\footnote{The Infrared Spectrometer And Array Camera mounted on the UT3 at the Very Large Telescope (VLT) and decommissioned in Period 92.} (SOFI) and were reduced using the PESSTO pipeline \citep[for further details see][]{sm14}. Multiple, dithered, on-source exposures were taken for \ca\ at each epoch; these images were flat-fielded and median-combined to create a sky frame. The sky frame was then subtracted from each of the individual images, which were then aligned and co-added. The NIR photometry obtained of the reference stars was calibrated against the Two Micron All Sky Survey (2MASS) catalog magnitudes \citep{sk06}. 
The NIR photometry for \ca\ are reported in Table~\ref{table:snir}.

Our optical and NIR photometric measurements were performed using a point-spread function (PSF) fitting technique, with a fit to the transient and the sequence stars using the {\sc snoopy}\footnote{{\sc snoopy} makes use of the photometry routines within the {\sc daophot} package. \url{http://sngroup.oapd.inaf.it/snoopy.html}} package within {\sc iraf}\footnote{{\sc iraf} is distributed by the National Optical Astronomy Observatory, which is operated by the Association of Universities for Research in Astronomy (AURA) under cooperative agreement with the National Science Foundation.}.

Since instruments with very different passbands were used for the follow-up of \ca\/ we applied a P-correction \citep{in15}, which is a passband correction similar to the S-correction \citep{st02,pi04}, and allows us to standardise photometry to a common system, which were Sloan for {\it griz} and Bessell for {\it UBVRI}. 
This procedure takes into account the filter transmission function and the quantum efficiency of the detector, and the intrinsic spectrum of \ca, but does {\it not} include the mirror reflectance, as this is relatively flat across the optical range. The P-correction also does not include a correction for atmospheric transmission, as this is accounted for when calibrating our magnitudes to the local sequence of tertiary photometric standards.
Our  spectroscopic coverage allowed us to compute the P-correction using only the observed \ca\/ spectra. As expected we found the largest effect in the {\it U}- and {\it g}- and $I$-/$i$-bands, with average corrections of $\Delta U\sim0.08$, $\Delta I=\Delta i\sim 0.10$ and $\Delta g\sim-0.18$~mag, respectively. Whilst we found an average correction of $\Delta B=\Delta V=\Delta R=\Delta r=\Delta z\sim0.01$. The greatest differences were found for the EFOSC2 filters (e.g. $\Delta g_{\rm (EFOSC2)}=-0.28$ mag and  $\Delta z_{\rm (EFOSC2)}=0.08$ mag). We note that the P-corrections from the $i\#705$ EFOSC2 filter to both $i$ Sloan and $I$ Bessell filters were of a similar magnitude, allowing us to transform our photometry to both of these systems. Unfiltered images of the SN (including the discovery epoch) were obtained with a camera which has a response peaking in {\it R}-band, and so these were calibrated to this bandpass.

{\it Swift}+UVOT data (in the {\it uvw2}, {\it uvm2}, {\it uvw1}, {\it u}, {\it b} and {\it v} bands) were reduced using the HEASARC\footnote{NASA High Energy Astrophysics Science Archive Research Center.} software package. Images obtained at the same epoch were co-added before aperture magnitudes were measured following the prescription of \citet{po08}.  A 3\arcsec\ aperture was used to maximise the signal-to-noise ratio (S/N).  The standard {\it Swift} zero points were applied and subsequently {\it Swift} {\it ubv} magnitudes were transformed to the Landolt system applying shifts of $\Delta U=0.26$, $\Delta B=0.02$ and $\Delta V=0.01$. These shifts were quantified from the magnitudes of the sequence stars in the SN field.

Longslit optical spectra were obtained with the NTT+EFOSC2 on La Silla and with Gemini South + Gemini Multi-Object Spectrographs (GMOS) on Cerro Pachon, both in Chile. Integral field spectra were taken with the ANU 2.3 m telescope + the Wide Field Spectrograph \citep[WiFeS,][]{do10} at Siding Spring Observatory in northern New South Wales, Australia.
All longslit spectra were reduced in the standard fashion (including trimming, overscan, bias correction, and flat-fielding) using standard routines within {\sc iraf}.
In the case of spectra observed with EFOSC2 these steps were performed within the PESSTO pipeline \citep{sm14}. Wavelength calibration was performed using spectra of comparison lamps acquired with the same configurations as the SN observations. EFOSC2 spectra were corrected for telluric absorption by subtracting a model spectrum of the sky bands.
For WiFeS the {\sc pywifes}\footnote{http://www.mso.anu.edu.au/pywifes/} package \citep{Chi14} was used to reduce the spectra and produce data cubes, from which the final spectra were obtained using a PSF weighted extraction routine. The resolutions of the longslit spectra were checked against the full width at half-maximum (FWHM) of narrow night sky emission lines and are reported in Tab.~\ref{table:sp}. Flux calibration was performed using spectrophotometric standard stars observed on the same nights and with the same configuration as 
\ca. The flux calibration was checked against the photometry by integrating the spectral flux under standard Sloan or Bessell filters, and adjusting by a multiplicative factor when necessary. The resulting flux calibration is accurate to within 0.1 mag. We note that at the wavelength of \Ha\/ EFOSC2 and GMOS spectra have a resolution in velocity space of $\sim700-800$ \kms\/, while WiFeS has a resolution $\sim90$ \kms.

NIR spectroscopy was obtained solely with NTT+SOFI. Spectra were wavelength calibrated via spectra of comparison arc lamps acquired with the same configuration as the SN observations. Solar analogs were also observed at a similar airmass and time to \ca\ in order to facilitate the removal of telluric absorptions between 1 and 2~$\upmu$m. For three epochs a flux standard was observed on the same night and using the same configuration as the SN spectra, and was used to calibrate these. For the other epochs we used the telluric standard and Hipparcos photometry to flux calibrate the spectra of \ca. The difference in measured flux between the two methods were within $0.2-0.3$ mag.

All PESSTO spectra are available through the ESO Science Archive Facility (SAF) as standard phase 3 ESO products.
The reduced images taken by PESSTO are also available for download from the PESSTO home page and data access instructions are available on \url{www.pessto.org}. All spectra, including the non-ESO data, are available on WISeREP\footnote{http://wiserep.weizmann.ac.il/} \citep{ya12}.

\section{Photometry}\label{sec:ph}

\subsection{Peak and explosion epoch}\label{sec:ep}
In this work we revised the epoch of maximum light with respect that of \citet{in14}. The previous peak epoch was based on a spectroscopic comparison of our classification spectrum with that of SN~1997cy, which does not have precise information about peak epoch. In this paper we present additional photometric points in the UV and in the optical (cfr. Tabs.~\ref{table:snm},~\ref{table:sns} and~\ref{table:snuv}) around the epoch of our first spectrum that allow us to better determine the peak epoch. Using a low order polynomial we fit the $R$-band data and found the maximum light to be at MJD $56048.2\pm4.4$ (2012 May 1.2), which is 60 days later than that previously reported.  
To estimate the uncertainty on the epoch of the lightcurve peak, we calculated the difference when using polynomial fits of second, third and fourth order; the difference in the peak epoch when fitting the $R$-band and $g$-band data; and the additional uncertainty due to the conversion from unfiltered to $R$-band magnitudes. For the latter, we refitted the peak of the lightcurve using the lower and upper extrema of the 1$\sigma$ photometric error on the unfiltered data. We then added all of these in quadrature to determine the total uncertainty of $\pm4.4$~d.

If we consider that \ca\/ is probably already affected by CSM interaction in the earliest data, the peak epoch is somewhat uncertain. While the rise to peak is based on a limited number of pre-maximum measurements, taken with different filters, the spectral comparison also suggests that \ca\/ was classified around peak or at the latest few weeks after. We note that the evaluation of the peak epoch has the largest effect on the kinetic energy estimate (see Sections~\ref{sec:bol}~\&~\ref{sec:dis}) in that more energy would be required to power the light curve if the peak epoch is earlier than our estimate.
Estimating the explosion epoch is made harder by the fact that we have no information about the underlying SN ejecta - and hence about the type of explosion mechanism - and when the interaction starts to dominate the light curve evolution. In $R$-band we observe a rise time of 0.19 mag in $\sim$5 days. If we assume that the contribution to the luminosity from interaction does not change during the light curve rise, and that there is at least a 2 magnitude rise from the explosion to the peak, we can make a very rough estimate that the explosion happened on MJD $55998.2\pm20$ (2012 March 12).

\subsection{Light and colour curves}\label{sec:ph}
Pre-peak observations are available in {\it gR} bands, while $UBV$ and UV start at peak. We continued to observe the SN until it disappeared behind the Sun in late November 2012. We continued the follow-up from March 2013 until October 2013, when the SN was too faint to be detected with NTT. 

The decline post maximum light is steady in all bands until November 2012 (shown as S1 in Fig. \ref{fig:lc}), when SN 2012ca disappeared behind the Sun. The decline in {\it R} is 0.62 mag 100d$^{-1}$, which is similar to that shown by SNe 1997cy and 2005g (0.75 and 0.88 mag 100d$^{-1}$ respectively), but slower than that of SN 2002ic (1.66 mag 100d$^{-1}$, measured in {\it V}-band) and PTF11kx (3.30 mag 100d$^{-1}$). Curiously, the only two SNe that showed type Ia spectra around peak, namely SN~2002ic and PTF11kx, are those with the fastest decline, whilst the decline of the other SNe resembles more that of type IIn SNe such as SN~2010jl (0.74  mag 100d$^{-1}$).

When SN 2012ca was again observable in 2013 March, its decline rate had steepened, and went on to show two noticeable phases. The first phase (labelled S2 in Fig. \ref{fig:lc}), lasts from 2013 March to June, 322 to 409 days after peak, with a decline of  1.33 and 0.93 mag 100$d^{-1}$ in {\it R}- and {\it V}-band. There is a second, steeper decline afterwards ending at 531 days post peak with  2.4 and 3.3  mag 100$d^{-1}$ in {\it R} and {\it V}, respectively (labelled S3 in Fig. \ref{fig:lc}). Such a change in the decline may indicate the end of the main interaction that powered the luminosity of the SN since its first detection (see Section~\ref{sec:dis}). The steepness of the decline looks to increase progressively as strengthened by a non-detection of the SN at 560 days after peak and the absence of SN spectral feature in the last spectrum (see Section~\ref{sec:sp}) obtained 30 days before the non-detection.

The UV data also show a constant, slow decline in the first 30 days after maximum light. A similar behaviour is shown in the NIR bands during the 2012 follow-up campaign. However, in early 2013 we see a flattening of the light curve that lasts until 551 days in {\it K}-band, while the decline in {\it J} and {\it H} from 346 days onward is slower than that observed in the optical bands. We measured decline rates of 1.31 and 0.51 mag 100~d$^{-1}$ for {\it J}- and {\it H}-band, respectively. This behaviour would suggest ongoing dust formation as a consequence of the interaction of the SN ejecta with the H-rich CSM, even if we do not see a blue shift over time in the \Ha\/ profile (see Section~\ref{sec:sp}).

\begin{figure}
\includegraphics[width=\columnwidth]{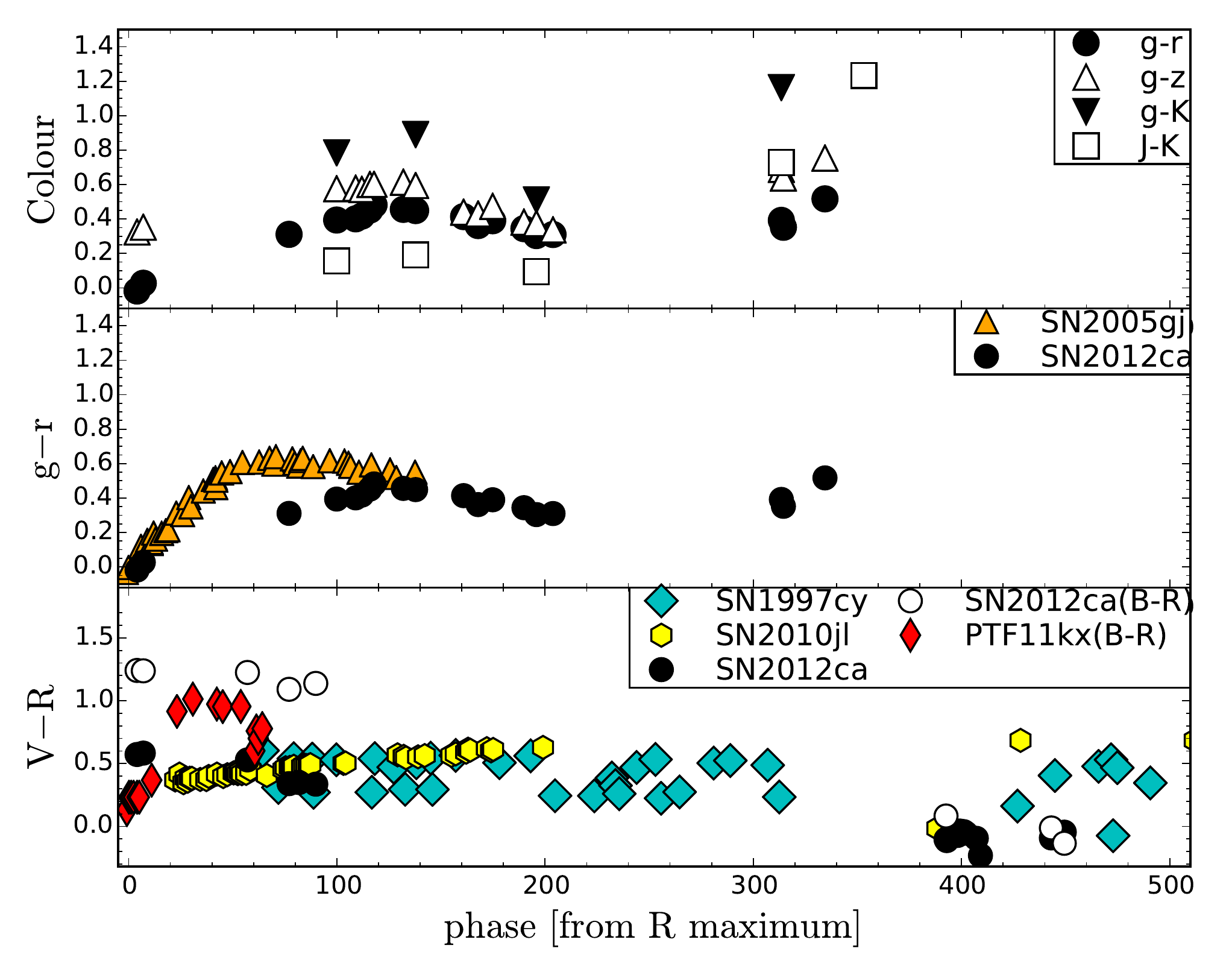}
\caption{ Top: dereddened colour evolutions of \ca. Middle: comparison of the dereddened $g-r$ colour curves of \ca\/ and SN~2005gj. Bottom: comparison of \ca\/ dereddened $V-R$ nad $B-R$ colour curves with that of PTF11kx ($B-R$) with those of SNe 1997cy and 2010jl (both $V-R$).}
\label{fig:col}
\end{figure}

The colour evolution of \ca\/ is shown in Figure~\ref{fig:col} together with that of SNe 1997cy, 2005gj, 2010jl and PTF11kx. We note the $g-r$ colour curve has a slow monotonic increase towards the red until $\sim$120 days. After that, it remains constant at $g-r\approx0.4$ until 320 days when it increases again toward the red. With the exception of $g-r$, all colours show almost constant evolution for the first 320 days. The last increase appears in all the colours and it is greatst in $g-K$ and $J-K$. The $V-R$ comparison shows \ca\/ has a constant optical colour through the entire evolution similar to that of SN~1997cy. Both SNe have a similar  behaviour to that of a prototypical SN dominated by interaction as SN~2010jl. The $B-R$ early increase in PTF11kx is due to the underlying thermonuclear ejecta.

\begin{figure*}
\includegraphics[width=18cm]{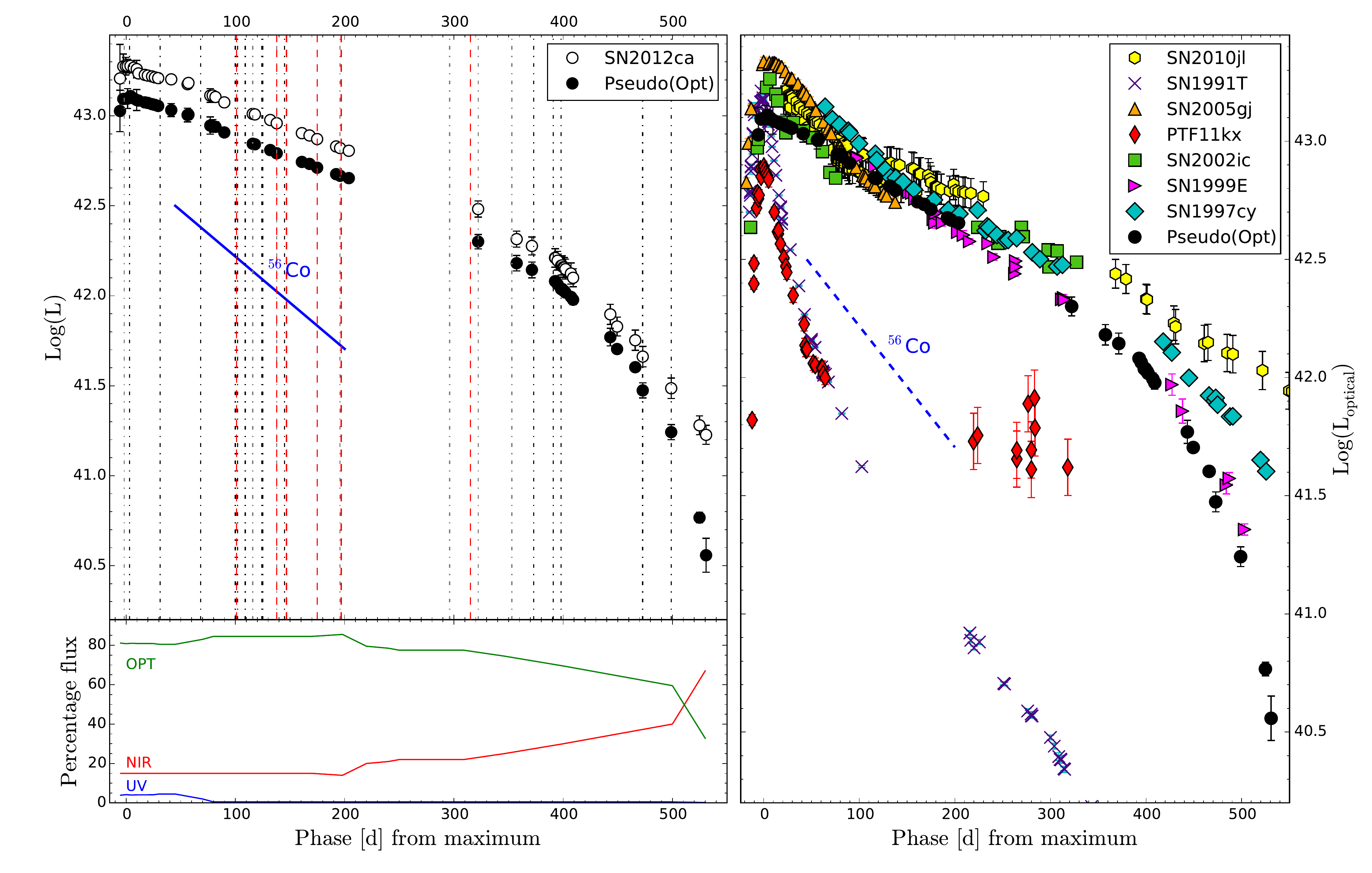}
\caption{Top left: pseudo ({\it U} to {\it z}, black points) and full (UV+OPT+NIR, white points) bolometric light-curves of \ca\/, as compared to the \co\/ decline which powers non-interacting SNe at late ($>$100d) phases. Dot-dashed black and grey lines indicate the epochs of optical spectroscopy presented here and in \citet{in14}, respectively. Red dashed lines indicate the epochs of NIR spectroscopy. Bottom left:  percentage of the bolometric flux at UV (blue), optical (green) and NIR (red) wavelengths for \ca. Right: comparison of \ca\/ pseudo bolometric lightcurve with those of other type IIn/Ia-CSM, the bright Ia SN~1991T and a classical type IIn as SN~2010jl. The pseudo bolometric lightcurves of comparison SNe have been created following the prescription in Section~\ref{sec:bol}.}
\label{fig:bolom}
\end{figure*}

\section{Bolometric luminosity}\label{sec:bol}

To obtain a direct measurement of the bolometric luminosity of a SN, UV to NIR photometry covering all epochs is required. While this is
typically difficult to obtain at all epochs during a SN light-curve, for \ca\/ we have extensive coverage. Nevertheless,
the UV data end at $\sim$30 days after peak, and so we extrapolated the UV contribution in time until 80d after maximum, when the extrapolation reaches zero (see Figure~\ref{fig:bolom}). The extrapolation was chosen over a blackbody fit to the observed bands at each epoch, as the latter does not well reproduce the spectral energy distribution (SED) of a SN interacting with a circumstellar matter.

To construct the bolometric lightcurve, the broad band magnitudes in the available optical bands were converted into
fluxes at the effective filter wavelengths, then were corrected for
the adopted extinctions (cfr. Section~\ref{ss:dr}). An SED was
then computed over the wavelengths covered and the flux under the SED
was integrated assuming there was zero flux beyond the integration
limits.  Fluxes were then converted to luminosities using the
distance adopted previously.
In the following, we will use the term ``pseudo-bolometric light-curve'' to refer to a bolometric light-curve
determined using only the optical filters and ``full bolometric light-curve'' for a light-curve including UV through NIR contributions.
We initially determined the points on the pseudo-bolometric light-curve at epochs when simultaneous (or very close in time) {\it UBVRIgriz} data were available;
for later epochs ($>$ 340 days) the bolometric luminosity was computed where there was coverage in less than four filters.  Magnitudes from the missing bands were generally estimated by interpolating the light curves using low-order
polynomials between the nearest points in time.  For some points this interpolation was not possible and we used an extrapolation assuming constant colours from neighbouring epochs.

In Figure~\ref{fig:bolom} (top left) we show the difference between the \ca\/ pseudo-bolometric and full bolometric light-curves. The overall luminosity of the full bolometric light-curves is obviously higher over time. 
The full bolometric light-curve peaks at $\mathrm {L_{\rm bol}}\approx1.90\times10^{43}$ erg s$^{-1}$, while the pseudo-bolometric light-curve has a maximum at $\mathrm
{L_{\rm pseudo-bol}}\approx1.29\times10^{43}$ erg s$^{-1}$. As shown in the bottom left panel of Figure~\ref{fig:bolom}, the NIR contribution increases from 200~d onwards, and becomes the main contributor to the total luminosity after 500 days. This NIR excess is an indication of dust formation. Dust formation has been seen in SN 2012ca-like objects, especially at mid-infrared wavelengths \citep{fox13}.


In the {\it Swift}+XRT data taken along with the first  seven epochs of UVOT imaging, which had average exposure times of $<$~2 ks, we did not detected any x-ray emission at the position of the SN. We set an upper limit to the x-ray flux of 8.3 $\times10^{-14}$ erg cm$^{-2}$ s$^{-1}$, corresponding to $L_{\rm X}<6.4\times10^{40}$~erg~s$^{-1}$. While this limit would have allowed us to detect the x-ray bright Type IIn SN~2010jl, which had a peak x-ray luminosity of $8.5\times10^{41}$~erg~s$^{-1}$ \citep{cha15}, many other SNe have x-ray fluxes below this limit \citep{dw12}.

In the right panel of Figure~\ref{fig:bolom} we show a comparison of \ca\/ pseudo-bolometric light-curve with those of other type IIn/Ia-CSM, namely SNe 1997cy \citep[{\it BVRI};][]{ge00,tu00}, 1999E \citep[{\it UBVRI}][]{ri03}, 2002ic \citep[{\it BVI};][]{ha03,wv04}, 2005gj \citep[{\it ugriz};][]{pr07} and PTF11kx \citep[{\it gri};][]{di12,si13b} together with a normal, long-lasting type IIn SN~2010jl \citep[{\it UBVRI};][]{zh12,fran14} and the bright type Ia SN~1991T \citep[{\it UBVRI};][]{li98}. Spectroscopic and photometric similitudes between SN~2002ic and PTF11kx - the only two objects showing early spectra {not dominated} by interaction - and bright type Ia SNe such as SN~1991T and SN~1999aa have already been shown in the literature \citep{ha03,di12}. 
\ca\/ shows a slowly declining light-curve that is inconsistent with the decay of \co\/ to \fe. A similar decline is seen in \ca, SN~1997cy and SN~1999E from 100d to 300d, while SN~2002ic and PTF11kx are also consistent from 200d to 300d  when their light curves are also dominated by interaction.
We note that the set of IIn/Ia-CSM SNe shown in Fig. \ref{fig:bolom} appear to divide into two classes. The first set of objects, comprising SNe 2012ca, 2005gj 2002ic, 1999E and 1997cy have a slow decline over timescales of $\sim$1 yr. Distinct from these is PTF11kx, which appears to show a similar peak luminosity, but a much more rapid initial decline similar to that seen in the Type Ia SN 1991T, followed by a slowly declining (and presumably interaction dominated) tail phase which is a factor $\sim$10 fainter than the first set of SNe.

\begin{figure}
\includegraphics[width=\columnwidth]{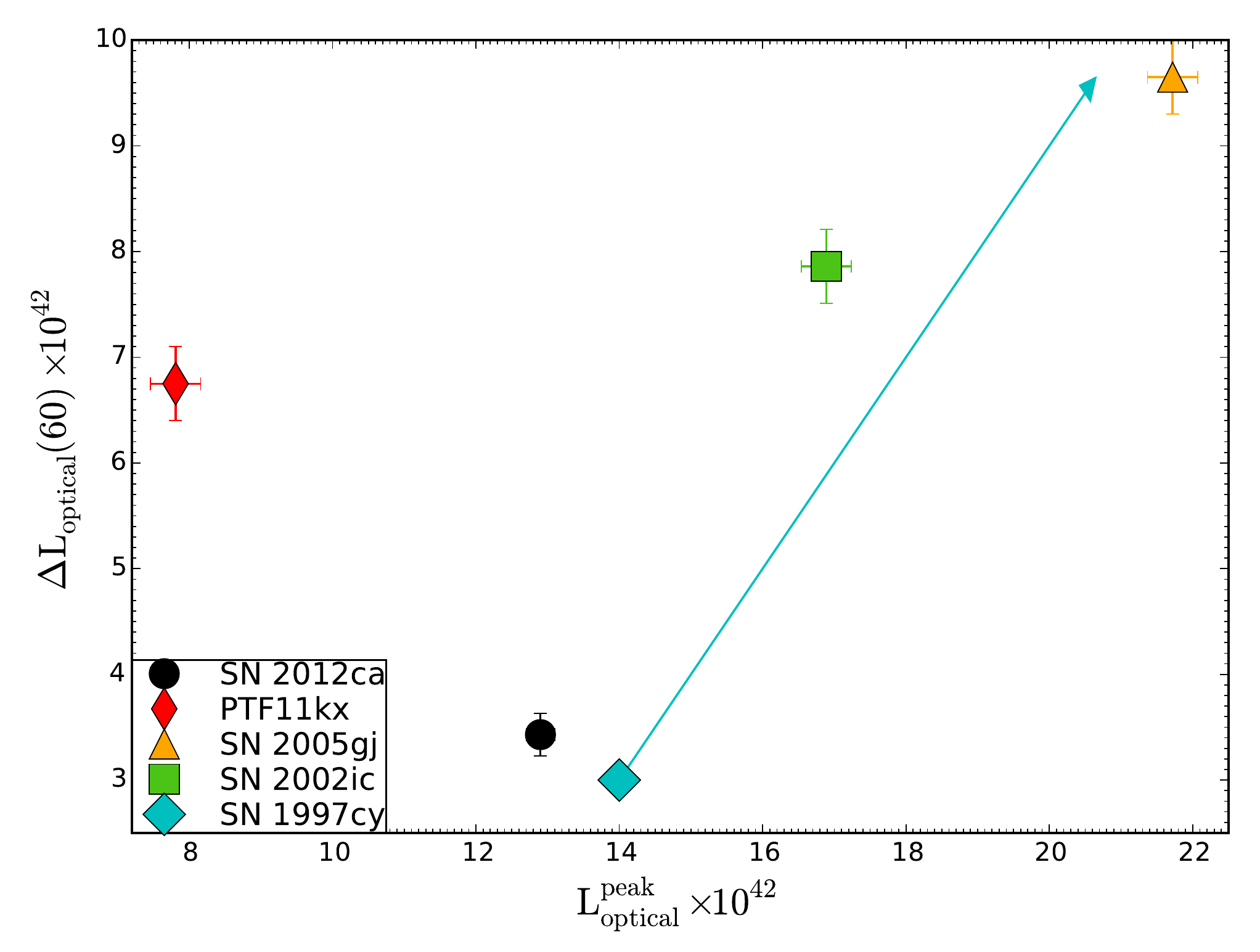}
\caption{Peak luminosity of pseudo-bolometric light curves of type IIn/Ia-CSM versus the luminosity difference from peak to 60 days (see text for more details). The arrow shown for SN~1997cy shows the range of possible peak luminosities and $\Delta$L. The 60 d cut-off has been chosen based on the available data for PTF11kx.}
\label{fig:dec}
\end{figure}

Bearing in mind the caveat regarding the uncertainties on the epoch of maximum light (see Section~\ref{sec:ep}), we note that the shape of the light-curve of \ca\/ around peak appears different to that of SN~2002ic and PTF11kx, with a flatter decline and a less rounded shape. From the other objects with coverage around peak, it appears that SNe with a higher peak luminosity have a slower decline post peak (see Fig.~\ref{fig:dec}). If the characteristics of the inner ejecta are the same for these SNe, then this could be explained with an increase in interaction due to a more massive or more dense CSM. \ca\/ does not appear to follow such a trend (although we caution that if the peak was earlier, it would also have been brighter), while SN~1997cy has insufficient photometric coverage around maximum to test its evolution.

After 150d the interaction is the main power source of the luminosity both for PTF11kx and the bright type IIn/Ia-CSM like \ca. However, the difference in luminosity at 150d between PTF11kx and a SN 1991T-like object is a factor 3, while for \ca\/ and the other comparison objects it is a factor 25. At one year post-maximum, \ca\/ and the other type IIn/Ia-CSM SNe are more than 100 times brighter than a SN 1991T-like SN Ia at a comparable epoch. This means that at +1 yr, a spectrum of \ca\/, or another SN with similar luminosity such as SN~1997cy should be dominated by the CSM interaction, and that naively the lines arising from the SN ejecta should comprise less than 1 per cent of the emergent flux. 

\begin{figure*}
\includegraphics[width=18cm]{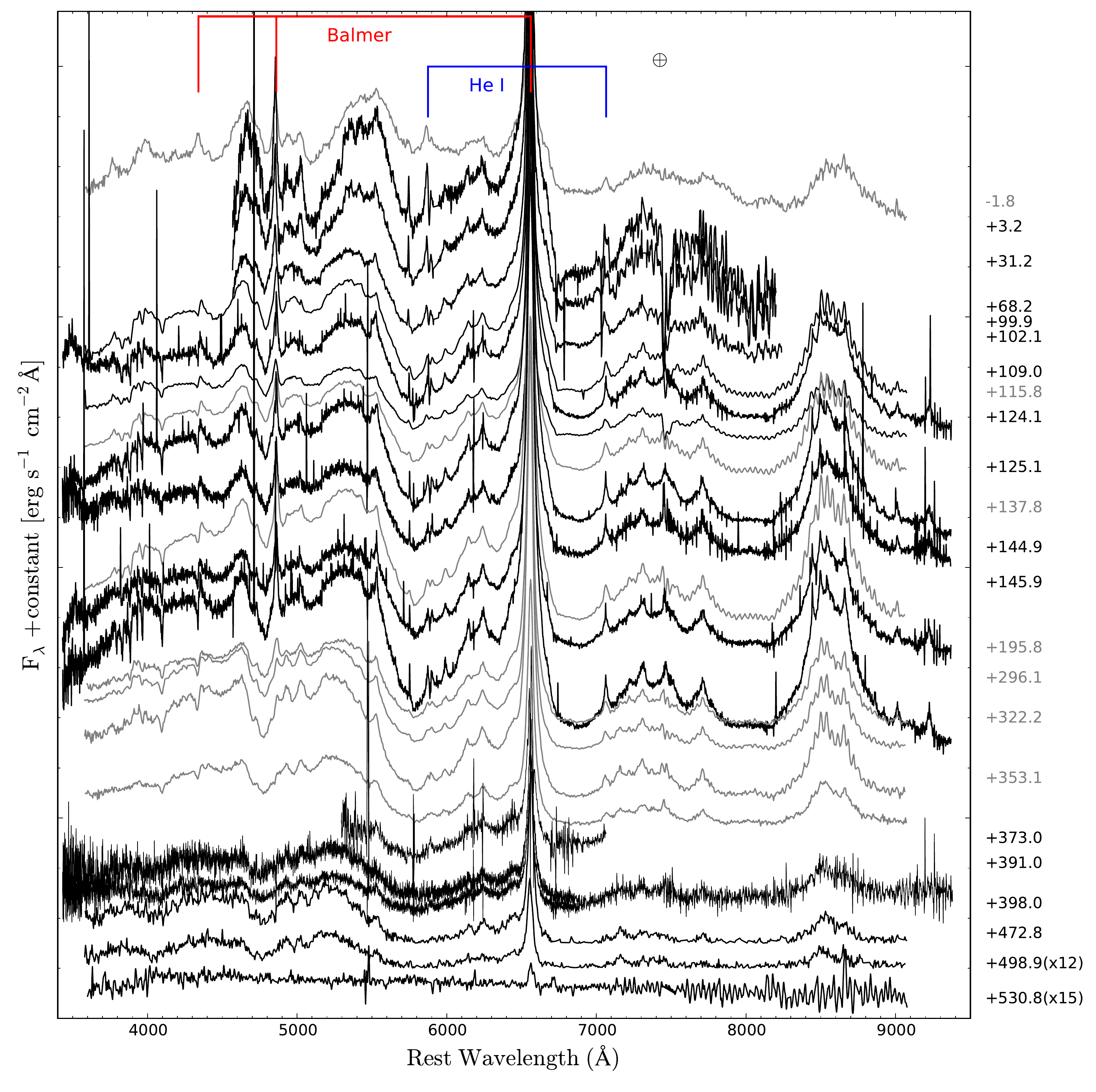}
\caption{Spectral evolution of \ca. The phase of each spectrum relative to the {\it R}-band maximum of the light curve is shown to the right of each spectrum. The spectra are corrected for Galactic extinction and are shown in the rest frame of \ca. The $\oplus$ symbol marks the positions of the strongest Telluric absorptions. The Balmer lines from \Ha\/ to H$\gamma$ are marked with red vertical lines, while He~{\sc i} $\lambda$5876 and $\lambda$7065 are marked in blue. Spectra in grey were already shown in \citet{in14}.}
\label{fig:spev}
\end{figure*}

\section{Spectroscopic evolution}\label{sec:sp}

In Figure~\ref{fig:spev} we show the optical spectroscopic evolution of \ca\/ from soon before maximum light to $\sim$500 days after peak. At all epochs, the spectrum is characterised by a relatively flat continuum, and narrow emission lines of H and He with a Lorentzian profile. Ca NIR is also present, while Fe lines and are blended together to form a blue ``pseudo-continuum'' below $\sim$5500 \AA.

The only noticeable evolution in the first 100 days is the slight relative change in flux between He~{\sc i} $\lambda$5876 and the surrounding region of the spectrum. The blue pseudo-continuum \citep{sm09}, which mainly arrises due to multiple Fe lines is visible since the first spectrum and does not evolve for the whole first year.
At wavelengths bluer than 5600\AA\/ the only differences from 400d onwards are the weakening of the Balmer lines and the decreasing toward the blue of the red shoulder of the pseudo-continuum (cfr. Figure~\ref{fig:last}). He~{\sc i}  $\lambda$5876 and $\lambda$7065 are also present from the earliest to the last spectrum, along with the NIR He~{\sc i} lines $\lambda$10830 and $\lambda$20589. The NIR Ca is seen in the first spectrum, and is still visible at 500 days post peak, while the Ca H\&K lines could contribute to some of the line at 4000~\AA. The former, broad feature is due to electron scattering and the decrease in prominence of the wings over time is indicative of a reduction of the phenomenon \citep[see Figures 2 \& 3 of ][ for further line identifications]{in14}. In a previous paper \citep{in14} we identified the two emission lines on the blue side of \Ha\/ as [O~{\sc ii}] $\lambda\lambda$6300,6364, and tentatively identified the line around 7000~\AA\/ as O~{\sc i} $\lambda$7774 which is blueshifted by $\sim$2500 \kms. However, \citet{fox15} noted that such apparent blueshifted O lines are seen at similar velocity offset in other type IIn/Ia-CSM such as SNe 2013dn and 2008J, making this explanation unlikely. Hence, a more plausible identification for these lines is multiplet 42 of Fe~{\sc ii} $\lambda$6248 and [Fe~{\sc ii}] $\lambda$7720.

The almost complete lack of evolution of the spectra of \ca\/ in 500 days is highlighted in Figure~\ref{fig:last} where the spectra from pre peak, +100 days and the final two spectra obtained are compared. The most striking difference is the appearance of a new line in the latest spectra at 7155\AA, which is not visible prior to 350 days. In order to confirm the presence of this line, in the bottom panel of Figure~\ref{fig:last} we show the residual after subtracting the spectrum at 100 days from the final spectrum taken at 500 days. Both spectra were taken with the same telescope and instrumentation (see Table~\ref{table:sp}). Aside from some broad undulations due to a change in the strength of the pseudo-continuum and some weak residuals at the position of H$\alpha$ and Ca~{\sc ii}, the only feature is the line at 7155\AA,  which we identify as [Fe~{\sc ii}] $\lambda$7155 with $v_{\rm FWHM}\sim 3900$ \kms\/. This line is usually seen in the late time spectra of both core-collapse and thermonuclear SNe, however we believe the origin of this feature is in the CSM, as it has a similar FWHM to \Ha\/ at that epoch ($v_{\rm FWHM}\sim 3000$ \kms\/). The absence of [Ni~{\sc ii}] $\lambda$7378 and [Ca~{\sc ii}] $\lambda\lambda$7291,7234 which are usually observed at similar epoch together with [Fe~{\sc ii}] strengthens the case that this line has CSM origin. The almost complete lack of evolution is also visible in the other SNe similar to \ca\/ that have spectroscopic coverage spanning at least over 300 days from the first spectrum dominated by interaction. In Figure~\ref{fig:lastc} such SNe, namely SNe 1997cy, 1999E and 2005gj show similar broad undulation to those of \ca\/ - which is shown for comparison - although weaker in amplitude. This suggest as the lack of evolution is a common feature among these SNe.

The NIR spectroscopic evolution of  \ca\/ (see Figure~\ref{fig:nirev}) is mostly dominated by the Paschen and Brackett series of H, together with the aforementioned He~{\sc i} lines. O~{\sc i} $\lambda$1.129 $\upmu$m is also visible as well as Mg~{\sc i} $\lambda$1.183 $\upmu$m, which is possibly blended with [Fe~{\sc ii}] lines.  In case of normal recombination, the presence of O~{\sc i} $\lambda$1.129 $\upmu$m would imply the presence of a stronger O~{\sc i} $\lambda$7774 line, which is not observed in our data. The absence of this line suggests that O~{\sc i} $\lambda$1.129 $\upmu$m line can be observed because of a fluorescence effect, as the O~{\sc i} $\lambda$1025 2p$^4$($^3$P)-3d($^3$D$^0$) transition lies within the Ly$\beta$ Doppler core and the O~{\sc i} $\lambda$1.129 $\upmu$m 3p($^3$P)-3d($^3$D$^0$) transition is optically thin \citep[see][for further details]{je12}.

In summary, the spectroscopic evolution of \ca\/ is dominated exclusively by H, He, Fe, Ca, and O, together with some small contribution from other metals such as Mg. Lines from ionised species, except Fe~{\sc ii}, are absent. As already shown by \citet{fr15} such elements are easy to reproduce by an inner hydrogen zone of a core-collapse SN showing only line flux arising from material present in the envelope of the pre-explosion progenitor \citep[based on the model of][]{je12}.

{The last two spectra (473d and 499d) do not show any significant dimming of the red wing of \Ha\/ or blue-shift in the peak of \Ha\/, which could be associated with the attenuation of emission originating in the receding layers by dust. Blue-shifted line peaks are one of the observational signatures of dust in SN ejecta and so this may appear in contrast to the NIR excess shown in Section~\ref{sec:bol}. However, this could mean that dust is not formed inside the ejecta but in a cool dense shell created by the SN--CSM interaction, close to the layer responsible of the hydrogen emission. Unfortunately, the absence of mid infrared observations at late times makes the determination of the site of dust formation very difficult. 

\begin{figure}
\includegraphics[width=\columnwidth]{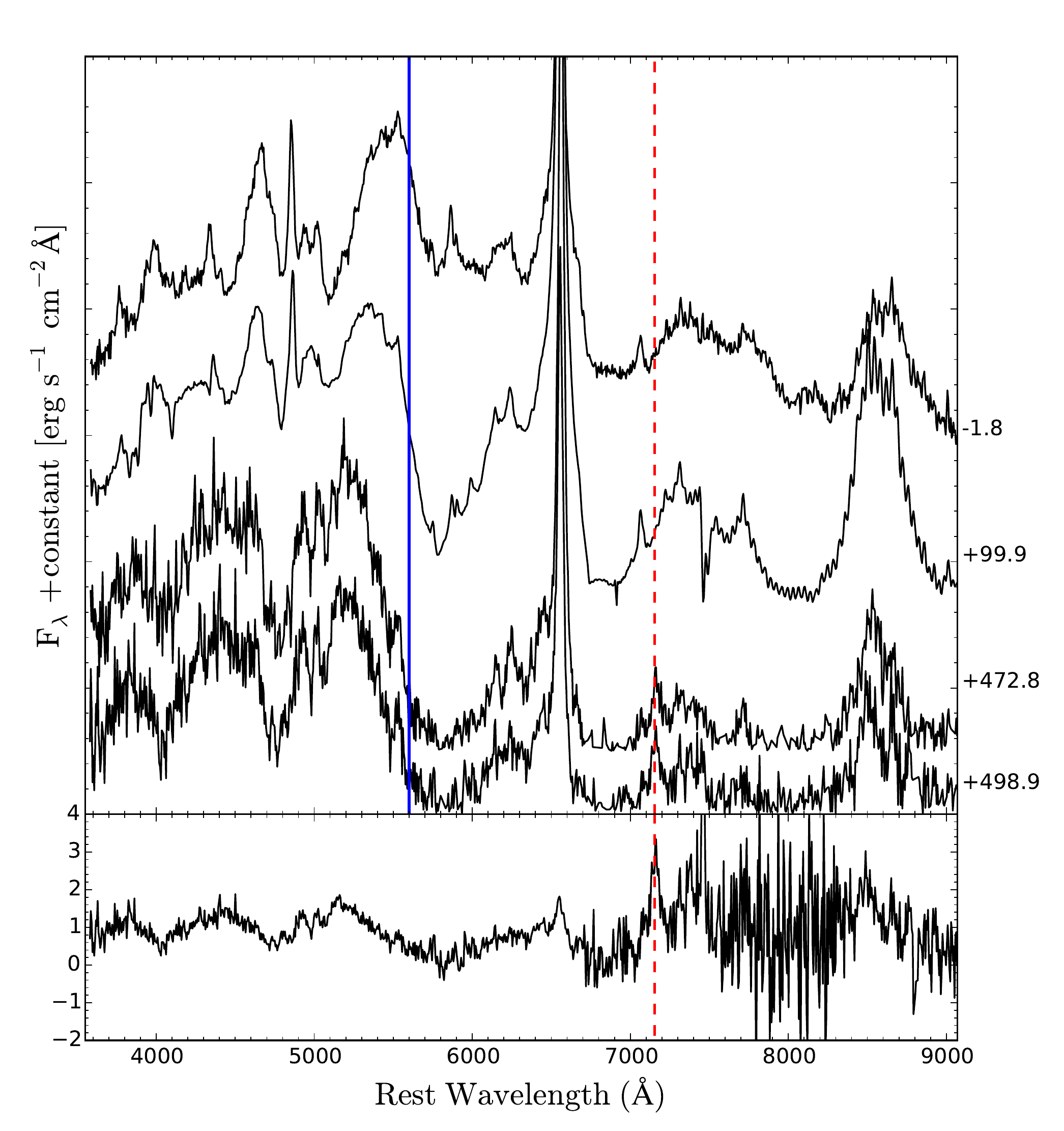}
\caption{Top: -2~d and 100~d spectra of \ca\/ compared to the final two spectra obtained (at +473 and +499 days from {\it R}-band maximum), to emphasise the lack of line evolution over 500 days. The blue solid line marks the blue edge of the pseudo-continuum. Bottom: Residual of the final spectrum at 499~d, after scaling to match the flux level  in the pseudo-continuum of the 100~d spectrum and subtracting it. The red dashed line marks the position of the [Fe~{\sc ii}] $\lambda$7155. line} 
\label{fig:last}
\end{figure}

\begin{figure}
\includegraphics[width=\columnwidth]{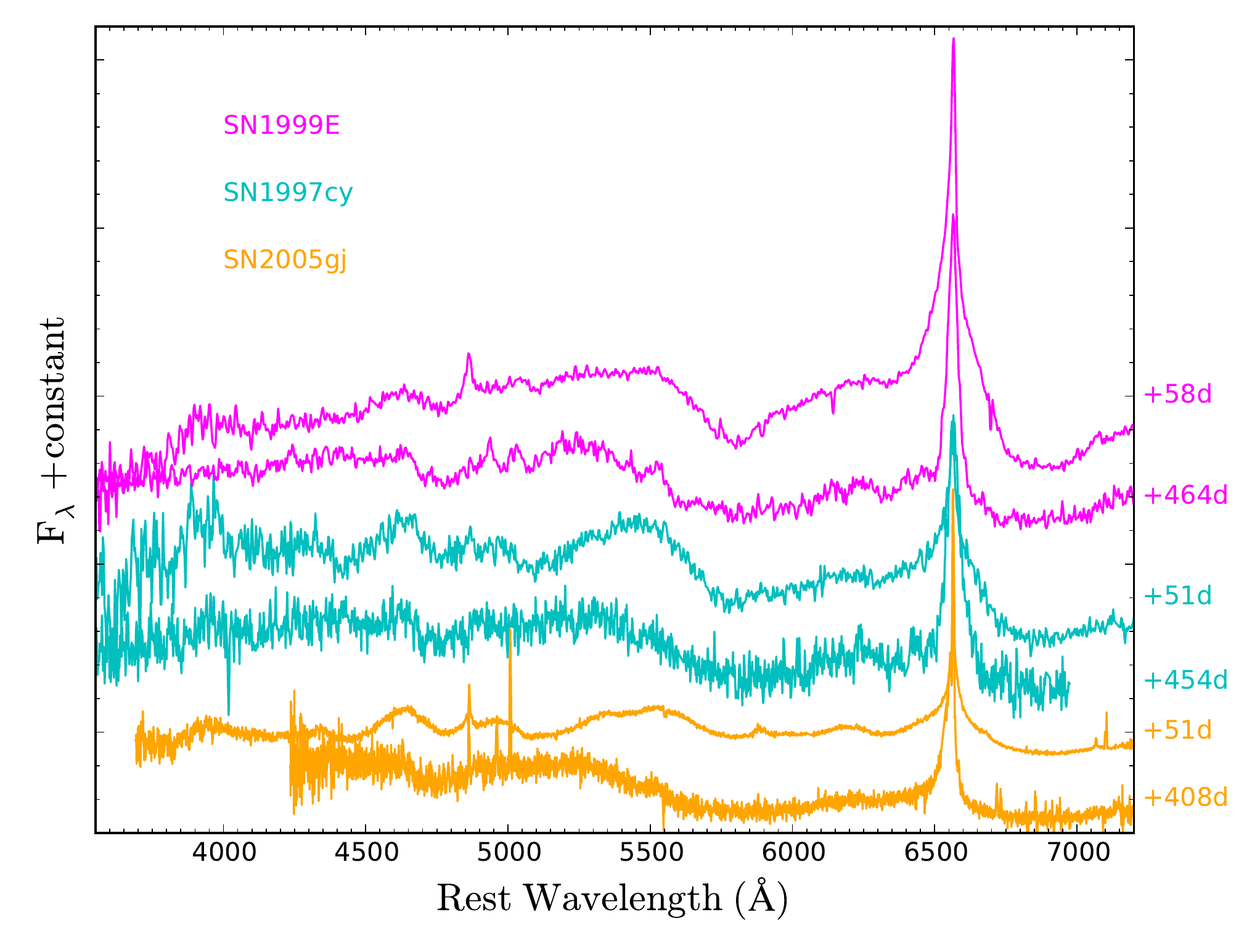}
\caption{Late time ($>$400d) spectra of SNe 1997cy, 1999E and 2005gj.} 
\label{fig:lastc}
\end{figure}

\begin{table*}
\caption{Main properties of type IIn/Ia-CSM}
\begin{center}
\begin{tabular}{lcccccc}
\hline
\hline
SN & Host type & Peak & Pre-interaction-dominated & Unshocked wind & {\it I}-band & References \\
 &    & (mag) & spectra & (km/s) &  double peak &\\
\hline
\ca 			&  late-type spiral 	& $-19.9$ ({\it R}) 		& No	 				& 200	& - 		&  (1) \\
SN 1997cy 	& compact-faint 	&$\leq-20.1$ ({\it V}) 	& No 				& - 		& - 		& (2) \\
SN 1999E 	& late-type irregular 	& $\leq-20.0$ ({\it V}) 	& No 				& 200 	& - 		& (3) \\
\multirow{2}{*}{SN~2002ic} 	& \multirow{2}{*}{compact-faint}  	&\multirow{2}{*}{$-20.2$ ({\it V})} 		& Yes (SN 1991T-like Ia 	& \multirow{2}{*}{-}		& \multirow{2}{*}{No}		& \multirow{2}{*}{(4)}	\\
			& 				& 				& or SN 2004aw-like Ic) 	& 	& 	& \\
SN 2005gj 	& late-type irregular  	&$-20.4$ ({\it r})		& No 				& 100-300	& No 	& (5)\\
SN 2008J 		& late-type spiral  	& $-20.3$ ({\it V})		& No 				& - 		& No 	& (6)\\
PTF11kx 		& late-type spiral   	& $-19.3$ ({\it R})		& Yes (SN 1991T-like Ia)	& 65 		& Yes 	& (7) \\
\hline
\end{tabular}
\end{center}
(1) - \citet{in14} and this work; (2) - \citet{tu00}, (3) - \citet{ri03}; (4) - \citet{ha03,de04,wv04,be06}; (5) \citet{al06,pr07,tr08}; (6) - \citet{ta12}; (7) - \citet{di12,si13b}
\label{table:sum}
\end{table*}

\begin{figure*}
\includegraphics[width=2\columnwidth]{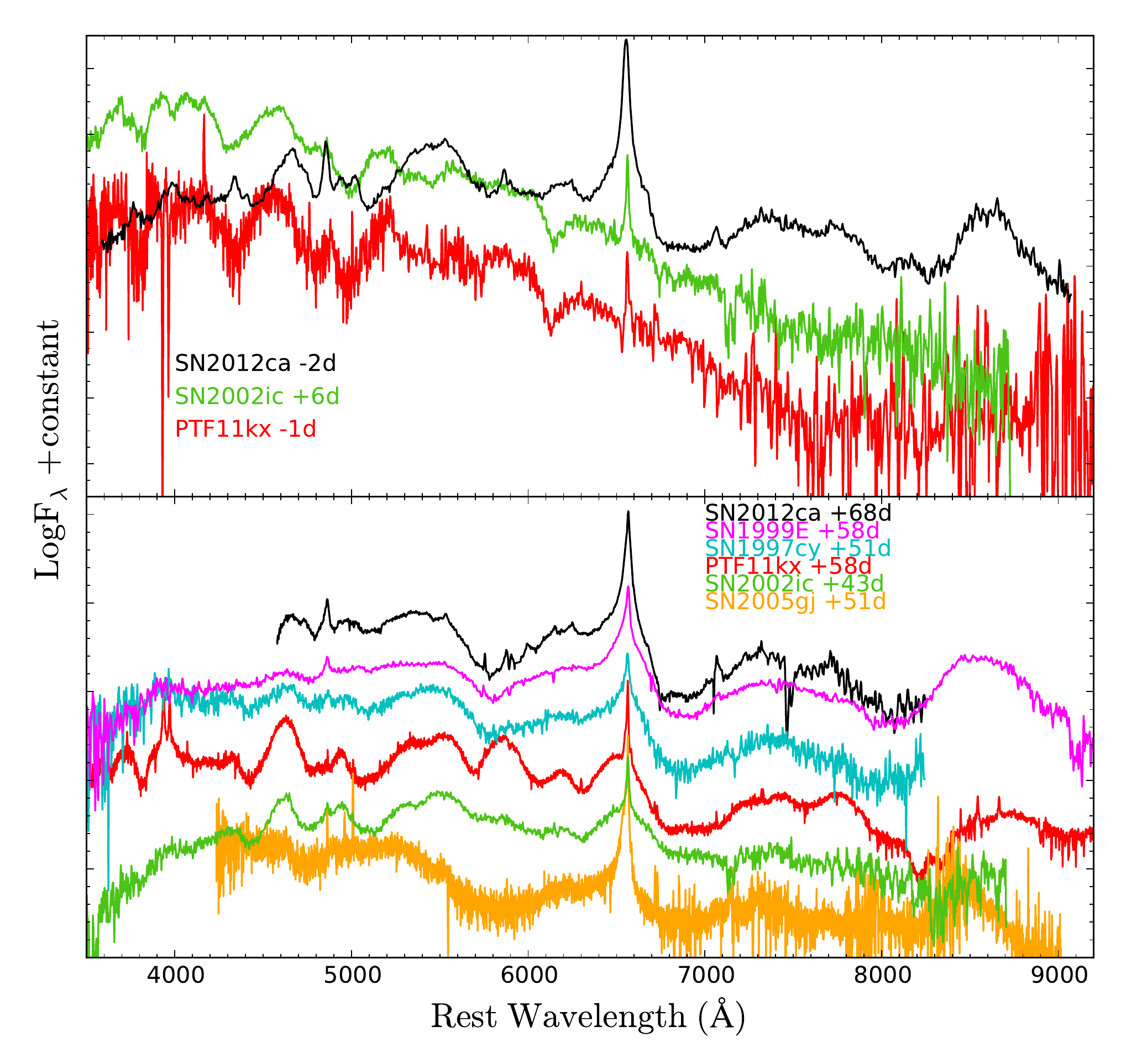}
\caption{Top: comparison of the spectra around the peak epoch of \ca\/, SN~2002ic and PTF11kx. The \ca\/ spectrum is dominated by interaction, while the other two SNe show the ejecta features together with narrow \Ha\/ due to the unshocked, H-rich CSM. Bottom: comparison of \ca\/ with other similar interacting SNe around 40-60 days after maximum light. We note as the phase of SNe 1997cy and 1999E have been evaluated assuming a rise time of 25 days from explosion to peak epoch.} 
\label{fig:spcmpt}
\end{figure*}

With the caveat that the epoch of maximum light for is not well constrained, the first spectra obtained for \ca\/ are different from those of SN 2002ic and PTF11kx. Indeed, as shown in the top panel of Figure~\ref{fig:spcmpt}, the spectrum of \ca\/ is already dominated around maximum light by H, Fe and Ca features arising from the ejecta interaction with the H-rich CSM. On the other hand (and with the exception of PTF11kx, which has the lowest degree of interaction as suggested from its photometric evolution), as the relative contribution of interaction to the lightcurve increases, the spectra of all objects in this observational class become more similar. However, some differences do remain: in the bottom panel of Fig.~\ref{fig:spcmpt} we see that \ca\/, SN~1997cy and SN~1999E show a similar multi-component H$\alpha$ profile with Lorentzian wings for the intermediate velocity component. Such wings are instead absent in both PTF11kx and SN~2002ic, which display a more Gaussian profile (for PTF11kx), or a combination of Gaussian and Lorentzian profiles (for SN~2002ic). The features arising in the blue pseudo-continuum and the NIR Ca~{\sc ii} triplet are comparable among  \ca\/, SN~1997cy and SN~1999E. The features bluer than 5000~\AA\/ exhibited by SN~2002ic and PTF11kx are stronger with respect to the pseudo-continuum, but this could be a consequence of a smaller contribution from interaction, with PTF11kx showing the weakest interaction among this sample.
The SN~2005gj spectrum has lower S/N than the others, but overall resembles SNe 1997cy, 1999E and 2012ca more than PTF11kx or SN 2002ic.

\section{On the origin of SN~2012ca and type IIn/Ia-CSM}\label{sec:dis}

\ca\/ is the best sampled object within the observational class\footnote{We use the term ``observational class'' here to identify a group of objects which share a set of broadly similar observational characteristics, while remaining agonistic as to whether these come from a single set of physically similar progenitors.} of possible type IIn/Ia-CSM SNe, albeit with limited information prior to maximum light. Moreover, it is the only object with UV monitoring. As a consequence, it can serve as an archetype for the post-maximum evolution of the set of type IIn/Ia-CSM SNe, or at least those which show a similar light curve, i.e. excluding PTF11kx.
As mentioned in Section~\ref{sec:bol}, the luminosity of \ca\/ at 365 days from peak, as well as the luminosity of SNe 1997cy, 1999E, 2002ic and 2005gj are more than 100 times brighter than a bright type Ia SN at a comparable epoch. This implies that the interaction of ejecta with CSM is responsible for $\sim$99 per cent of the total luminosity of the SN at late times.
Moreover, only the CSM should be responsible for the line flux observed in \ca\/ from 365 days onward.  We note that spectropolarimetric information is available for SN~2002ic at $\sim1$ year from maximum \citep{wa02} and suggests a dense, clumpy, disk-like CSM. Hence at early epochs there could be some caveats about the ejecta flux contribution due to viewing angle. Since we have shown that there is no spectroscopic evolution in \ca\/ from the epoch of peak luminosity onwards, and that the relative intensity between lines does not change abruptly or significantly, then we can assume that the spectroscopic evolution of \ca\/ is a consequence only of the physical conditions within the CSM and is not related to the underlying SN ejecta at any epoch.
A similar result can be obtained by applying the criteria for a SN to be visible under circumstellar interaction presented by \cite{le15}. Taking the absolute magnitude of SN 2012ca at peak in V-band ($-19.3$), we find that for a SN to be effectively ''hidden'' under the interaction requires it to be fainter than $V<-17.7$.
The remarkable spectroscopic similitude between \ca\/ and SNe~1997cy and 1999E suggests that this result also applies to the latter two. In contrast, this is only true for SNe like SN~2002ic and PTF11kx after 60-100 days, and when the early non-dominated interaction spectra have disappeared. In general we can assert that all the objects classified as type IIn/Ia-CSM and having luminosity comparable to that of \ca\/, SN~1997cy, SN~2002ic and SN~2005gj after 150 days have spectra fully dominated by the interaction, since the ejecta contribution to the observed spectrum is $\leq$5\%.

\begin{figure*}
\includegraphics[width=18cm]{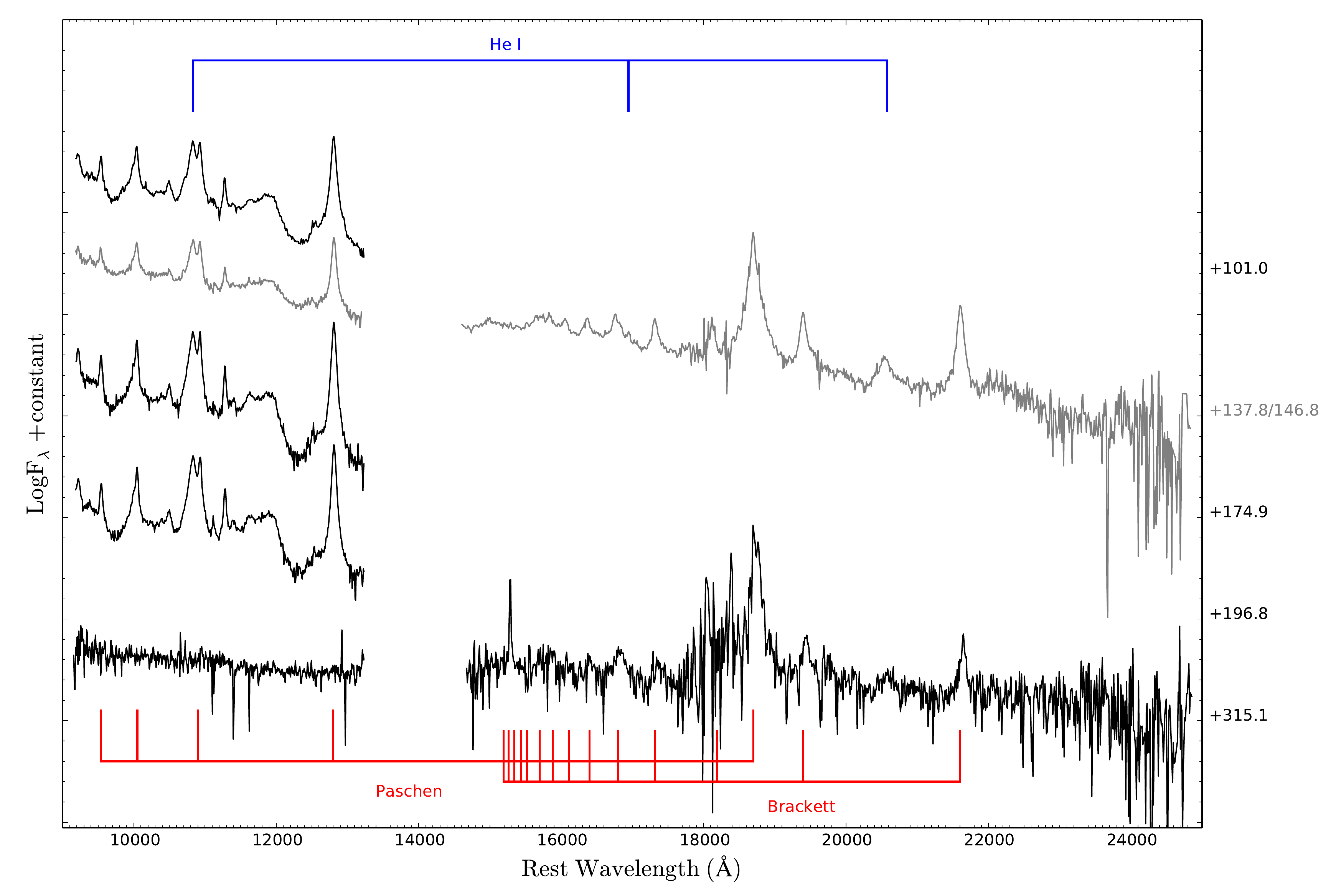}
\caption{NIR spectra evolution of \ca. The phase of each spectrum relative to light curve {\it R}-band maximum is shown on the right. The spectra are corrected for Galactic extinction and reported in the rest frames. Paschen and Brackett lines are marked with red vertical lines, while He~{\sc i} is shown in blue. A logarithmic scale has been chose to highlight the feature redwards of 14000~\AA\/. } 
\label{fig:nirev}
\end{figure*}

A physical constraint on the light-curves of interaction powered SNe is that their integrated luminosity from CSM interaction must be lower than the total kinetic energy available from the SN ejecta. For \ca\/ the integrated luminosity under the bolometric light-curve is E$=3.1\times 10^{50}$ erg, while the available energy found in typical type Ia SN explosion models is E$_{\rm k}=4.5-15 \times 10^{50}$ erg \citep{hn00,ro07,ma07}. If \ca\ is powered by the collision of the ejecta of a type Ia SN with a CSM, then it would necessitate a conversion efficiency from kinetic energy to luminosity of between 20 and 70 per cent. For comparison, \citet{fr13} estimated from the relative ejecta and shock velocitites that the conversion efficiency in the interacting transient SN 2009ip could be as high as 90 per cent. Furthermore, if the peak happened before earlier than that estimated (see the uncertainties in Section~\ref{sec:ep}), the integrated bolometric luminosity of \ca\ would be larger, implying a higher conversion efficiency, making the thermonuclear scenario even less likely. Core-collapse SNe can span from E$_{\rm k}=1 - 50 \times 10^{51}$ erg \citep{fi97,iw98}, and hence have a factor 2 to 100 times more energy available than from thermonuclear explosions. This would allow a lower, and perhaps more plausible conversion efficiency of between 1 and 30 per cent. However, while an efficient conversion may be required, a thermonuclear explosion cannot be ruled out for \ca\/ on the basis of energetics.

We also note that after a year from maximum light the bolometric light-curve for \ca\ changes slope as discussed in Section~\ref{sec:ph}. A change in the slopes of SNe 1997cy and 1999E was also observed at a similar phase. These three SNe are the only ones with data after a year from peak. The change in decline lasts almost 90 days for \ca\/ and SNe~1999E, and~1997cy. That could be a consequence of the fact that the reverse shock stops playing a role and the forward shock is slowing down rapidly. This happens when the mass of the shocked CSM is comparable to that of the ejecta and the reverse shock finishes crossing the ejecta \citep{svi12}. The slope of the decline depends on whether the shock is adiabatic ($n$=5, Sedov-Taylor phase)\footnote{$n$ is the index of the power-law describing the density profile of the moving ejecta.} or the internal energy behind the shock is radiated away ($n=3-4$, snowplow phase). A power-law decline with $n$=4 is similar to what is observed in the bolometric lightcurve of \ca\/ from 360 to 450 days and the pseudo bolometric lightcurve of SN~1997cy from 410 to 500 days (see Figure~\ref{fig:snp}). At these epochs the pseudo bolometric light curves of these objects have a steeper decline due to the decrease of the optical contribution to the total flux (see Section~\ref{sec:bol}). This is broadly consistent with that experienced by SN~2010jl \citep{of14}. The second, even more abrupt decline, could be due to an increase in dust production leading to additional attenuation of the optical flux, or because the forward shock became definitively inefficient  as a consequence of a drop in the wind density \citep{ch11}.

Another feature that could give us a valuable insight into the progenitor scenario is the narrow P-Cygni absorption seen in the line profile of \Ha\ (see right panel of Figure~\ref{fig:spcmp}). As discussed in \citet{in14}, in high resolution spectra of \ca\/ taken with WiFeS we observe a narrow P-Cygni absorption in \Ha\/ with $v\sim200$ \kms\/.  This velocity is a factor 2 or 3 greater than that
observed in probable diffuse CSM arising from the progenitor companion to some SNe Ia \citep[$50-100$ \kms;][]{pa11} and similar to those observed in Luminous Blue variable (LBV) winds , but is still comparable to that observed in a small fraction of SNe Ia \citep[$150-200$ \kms;][]{st11}. We also note that SN~1999E had a similar narrow P-Cygni absorption in \Ha\/ \citep[$v\sim200$ \kms\/;][]{ri03}, while the less luminous PTF11kx has a lower velocity absorption at $v\sim65$ \kms\/ \citep{di12}. Once again, the wind velocity is potentially consistent with both peculiar thermonuclear and core-collapse scenarios for \ca\ and other events such as SNe~1997cy and 1999E.

A possible additional clue is the variation of the line profile in the region 5800~\AA\/ -- 7200~\AA\/. As reported in Section~\ref{sec:sp} this region is the only one which evolves from the first spectrum before peak onwards. 
However, the sharp profile on the blue edge of this region, mainly dominated by \Ha\/, is not observed at any stage in PTF11kx and SN~2005gj but are shown by SN~1997cy and SN~1999E (cfr. Figure~\ref{fig:spcmp}).  SN~2002ic seems a borderline case but the overall shape is more similar to PTF11kx and SN~2005gj than the others as already observed in the comparison with SN~1997cy shown by \citet{ha03}. Although this is mere phenomenology, it is interesting how SNe 1997cy, 1999E and 2012ca are almost identical, and dissimilar to PTF11kx.

\begin{figure}
\includegraphics[width=\columnwidth]{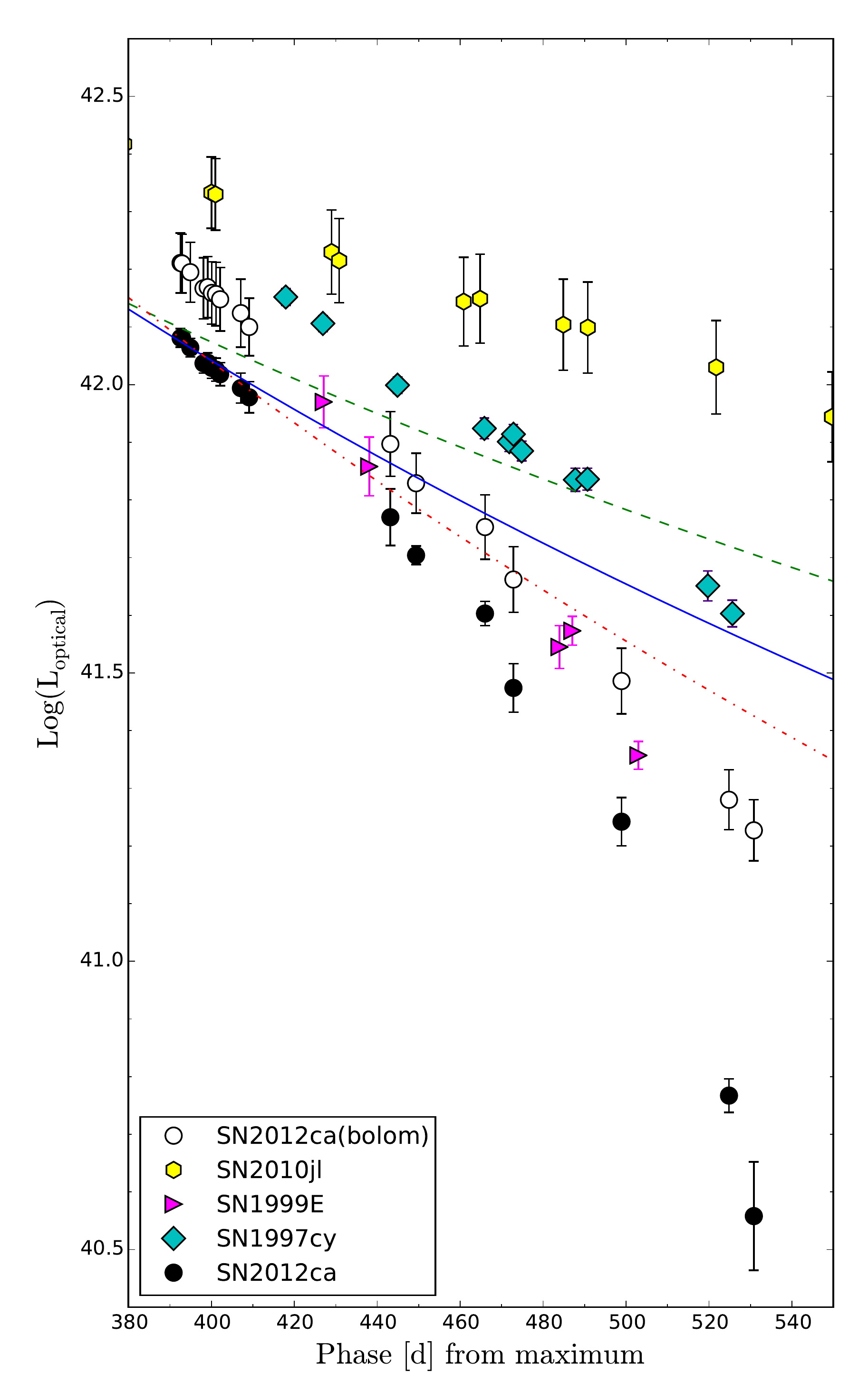}
\caption{The bolometric and pseudo-bolometric evolution of \ca\/, together with the pseudo bolometric lightcurves for SNe 1997cy, 1999E and 2010jl after one year. The blue solid (n = 4) and green dashed (n = 3) lines show the exponential decline for a snowplow phase, while the red dot-dashed line that of a Sedov-Taylor phase.} 
\label{fig:snp}
\end{figure}

\begin{figure}
\includegraphics[width=\columnwidth]{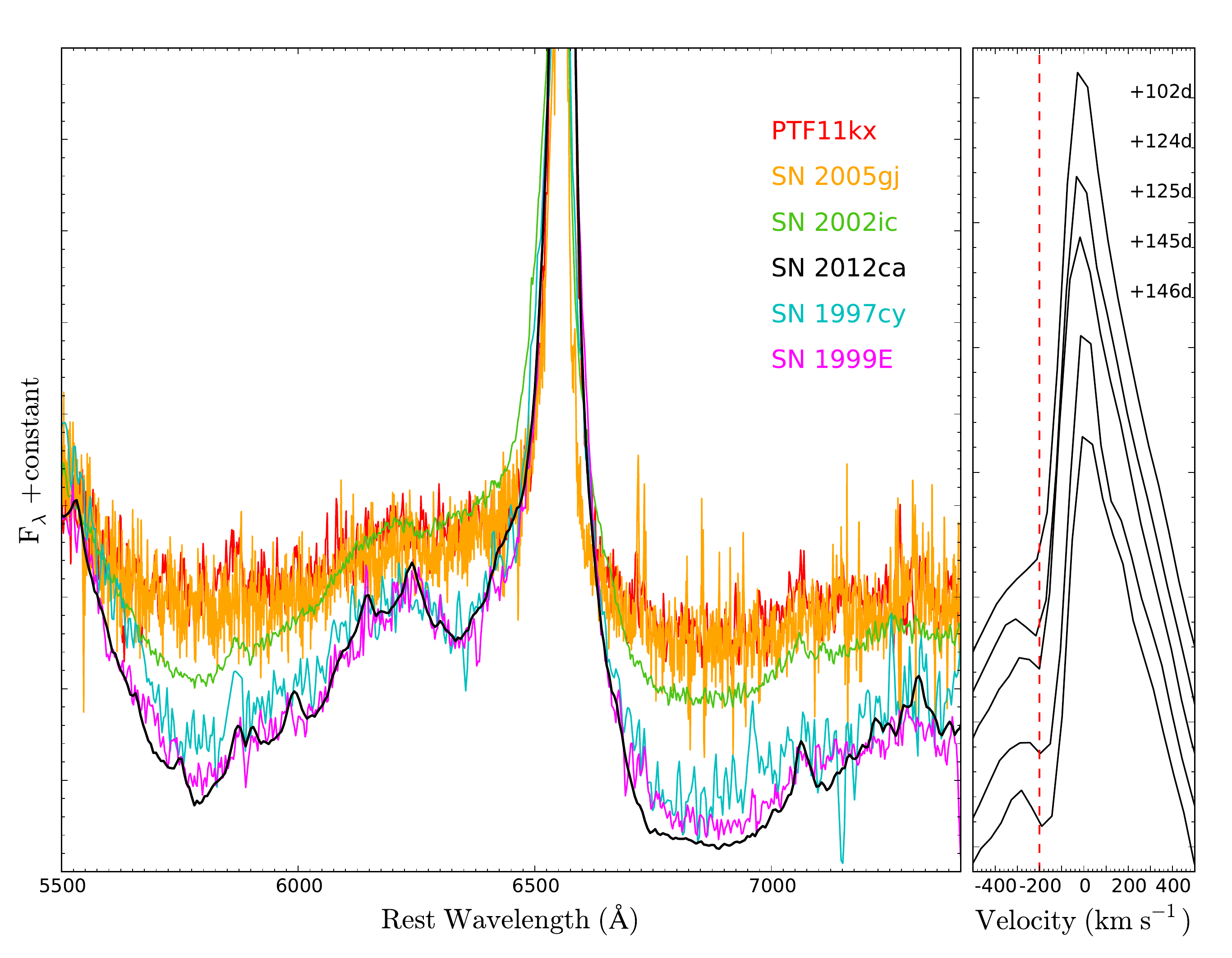}
\caption{Left: zoom in of the 5800~\AA\/ -- 7400~\AA\/ spectral region for \ca\/ and other type IIn/Ia-CSM, taken later than 100 days post-maximum, when the interaction can be seen to dominate their photometric evolution. Spectra are shifted in flux in order to match their blue pseudo-continuum. Right: temporal evolution of \Ha\/ from $\sim$100d to $\sim$150d post maximum. The red dashed line marks the velocity of the CSM.} 
\label{fig:spcmp}
\end{figure}

\subsection{Modelling the CSM of type IIn/Ia-CSM}

In Section~\ref{sec:sp} we have shown that all the elements observed in the spectra of \ca\/ can be explained as originating in the CSM. This, together with the lack of spectral evolution over 500 days suggest how little information on the underlying physical cause of these SNe can be obtained from their spectra. As an alternative approach, we compared simple models to the light-curves of these SNe in order to retrieve information on the ejecta and CSM masses.

We fit our light curves with an analytical model implemented by \citet{ni14} and based on the equations of \citet{cha12} that presented a semi-analytical model for the case of ejecta colliding with an optically thick CSM. The approximation used for the radiative diffusion of shock-deposited energy in the CSM is similar to that of \citet{ar89}. 
We note that for SNe Ia an exponential ejecta profile is perhaps preferred to a power-law. However, as already stated by \citet{wv04}, this profile does not yield an analytical solution and so we assume a power-law profile.

Since we have a H-rich CSM we set our opacity to be $\kappa = 0.34$ cm$^2$ g$^{-1}$, and allow five variables to vary freely in the fit, namely the ejecta, CSM and \ni\/ masses, the inner radius of the photosphere/pseudo-photosphere formed in the CSM, and the kinetic energy of the SN explosion. We also experimented with different CSM density profiles, including an $r^{-2}$ profile as expected from a stellar wind, and a uniform density shell which could be produced by a single eruption. Finally, we tested the effect of applying additional physically-motivated constraints on the possible ejecta and CSM masses. We limit the ejecta mass to that expected for a type Ia SN \citep[M$_{\rm ej}<1.5$\M,][]{sc14}, as well as that of \ni\/ \citep[M$_{\rm ej}<1.4$\M,][]{chi15} that is more than 20\% of the ejecta as suggested by \citet{un08}. We also constrain the CSM mass to less than the mass of a possible asymptotic giant branch (AGB) or super-AGB star companion \citep[M$_{\rm CSM}<8.0$\M;][and references therein]{mey15,pu09}, as stars more massive than this are expected to explode as core-collapse supernovae in a relatively short time ($\lesssim$40 Myr), and will hence are unlikely to survive long enough to produce the CSM for SN 2012ca if it arises from a Type Ia explosion.

The steady wind models give the lowest $\chi^2$. Fig.~\ref{fig:12cafit} shows the best fits with and without the aforementioned constraints on progenitor and ejecta mass. To reproduce the bolometric light-curve of \ca\/ we require M$_{\rm ej}=1.25$ \M\/ and M$_{\rm CSM}=2.63$ \M\/ (see Table~\ref{table:fit}). Although the ejecta mass is consistent with that of a type Ia SN, the mass of CSM is higher than that expected to be stripped from the companion \citep[0.11-0.20 \M\/;][]{pan10,pan12,liu12}. 
The kinetic energy retrieved E$_{\rm k}\sim7-9\times10^{51}$erg would need a conversion efficiency from 500\% to 2000\% in the case of thermonuclear origin. On the other hand, such a kinetic energy would be reasonable for energetic core-collapse events (e.g. a broad line type Ic 10\%-20\%) and would need a conversion efficiency of 50\%--70\% for normal type Ibc. However, we caution that the model used has several free parameters, and is a simplified approximation which ignores potentially important effects such as deviations from spherical symmetry, or a clumpy CSM. In the absence of more sophisticated modelling, the ranges of masses (see Table~\ref{table:fit}) seem more plausibly associated with a core-collapse origin for SN 2012ca.


\begin{table*}
\caption{Best-fit derived parameters for CSM modelling of the bolometric light curves and $\chi^2_{\rm red}$.
M$_{\rm ej}$, M$_{\rm CSM}$ and M$_{\rm Ni}$ $^{56}$ are the ejecta, CSM and ejected $^{56}$Ni mass respectively.  $R$ is the photospheric radius, E$_{\rm k}$ is the kinetic energy of the ejecta, and t$_{0}$ is the time from explosion to maximum light.
In the case of SN 2005gj, the best fitting model has a $^{56}$Ni mass which is larger than the ejecta mass. Since this is clearly unphysical, we also report the best fitting model where M$_{\rm Ni}$ is forced to be zero. We also report the parameters of the best fitting $^{56}$Ni-free model for SN 2012ca.}
\begin{center}
\begin{tabular}{lccccccc}
\hline
\hline
Object & t$_{0}$ & M$_{\rm ej}$ &M$_{\rm CSM}$ &M$_{\rm Ni}$ & R & E$_{\rm k}$  &$\chi^2$\\
& (days)   &(\M\/) &(\M\/)  &(\M\/) &($10^{14}$ cm)  &($10^{51}$ erg) &\\
\hline
\ca (with Ni) 			&  40.36 		& 0.93 		& 2.37  		& 0.52		& 5.09		& 6.82		& 1.95	\\
\ca (without Ni) 			&  34.57 		& 1.25 		& 2.63  		& 0.00		& 6.48		& 8.93		& 1.48	\\
SN~1997cy 			& -   			& 2.30		& 0.42 		& 0.85		& 6.57 		& 0.64		& 21.17		\\
SN~1999E 			&  -  			& 1.64 		& 2.93 		& 0.90 		& 4.56		& 0.06		& 8.06		\\
SN~2002ic 			&  -  			& 0.60		& 4.35 		& 0.00		& 6.14 		& 1.01		& 4.05		\\
SN~2005gj (with Ni) 		&  19.16 		& 0.82		& 1.70		& 1.06		& 5.53 		& 0.93		& 2.64	\\
SN~2005gj (without Ni)	&  18.37 		& 0.40		& 3.80		& 0.00 		& 1.10 		& 1.54		& 13.74	\\
PTF11kx 				&  -  			& 1.34 		& 0.07 		& 0.00		& 9.10		& 0.67		& 0.48		\\
\hline
\end{tabular}
\end{center}
\label{table:fit}
\end{table*}%

\begin{figure*}
\includegraphics[width=2\columnwidth]{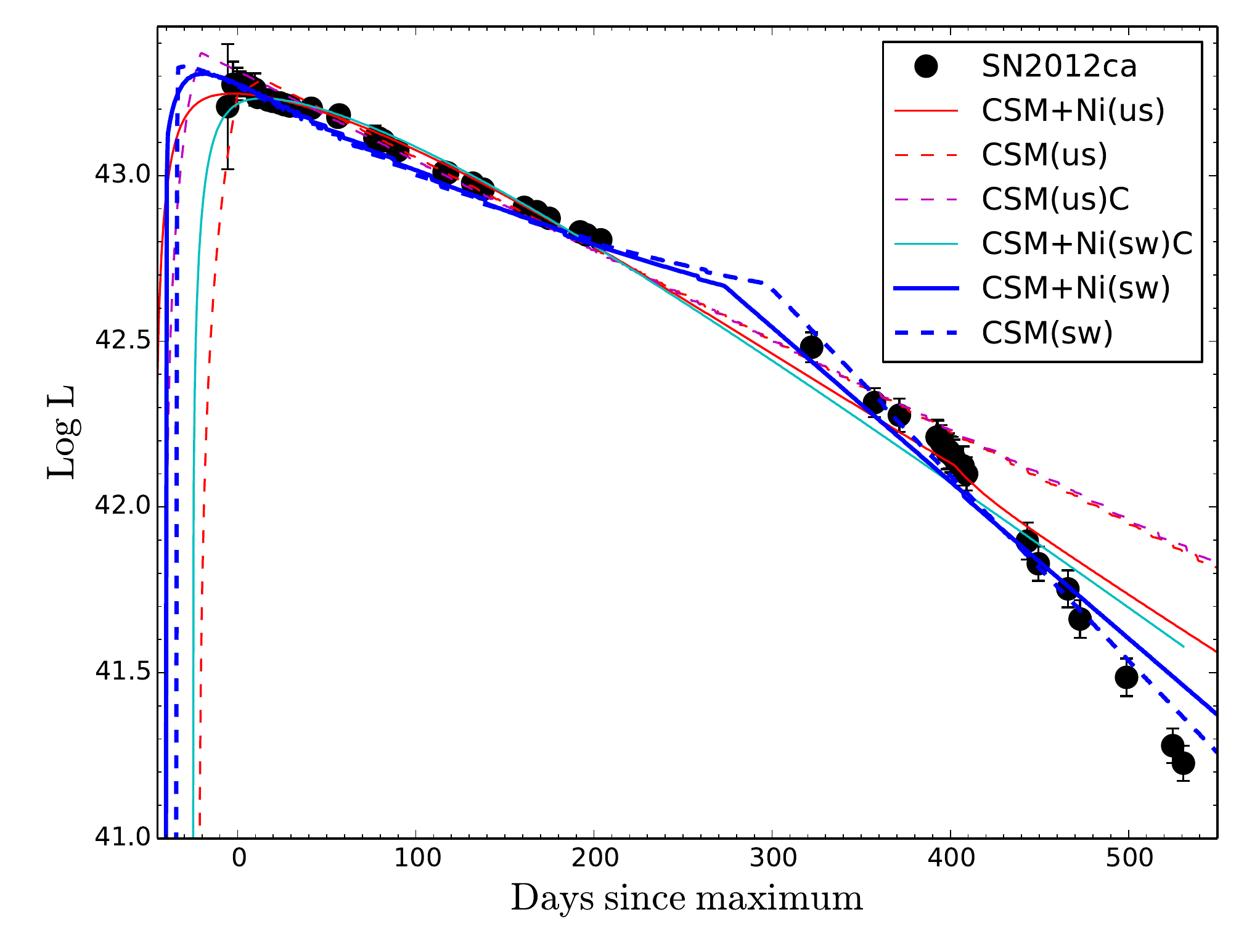}
\caption{The bolometric light-curve of \ca\/ and the semi-analytical models that best fit the light curve with and without the constraints mentioned in the main text. The fits with the lowest $\chi^2$  are the model with a CSM formed through a steady wind, $^{56}$Ni and no physical constraints ($\chi^2$ = 1.95; blue solid line) and the model with a CSM formed through a steady wind, no $^{56}$Ni and no physical constraints ($\chi^2$ = 1.45; blue dashed line). The suffix C refers to models with constraints (see text). ``SW'' and ``US'' refer to models with a steady wind and a uniform shell respectively.} 
\label{fig:12cafit}
\end{figure*}

While \ca\/ is the only type IIn/Ia-CSM with a full UVOIR bolometric light-curve, we repeated the previous analysis on the sample of similar SNe with pseudo-bolometric light-curves from Sect.~\ref{sec:bol}. As a consequence the parameters retrieved have to be considered as limits, especially for the kinetic energy. The best fit and related parameters are shown in Figure~\ref{fig:allfit} and reported in Table~\ref{table:fit}. 

We note that both the ejecta and CSM mass for PTF11kx appear consistent with a thermonuclear scenario, which is encouraging. Conversely, we were unable to fit the section of the light-curve for SN~2002ic which is affected by interaction with a combination of masses appropriate for a thermonuclear explosion, even assuming that the companion is a massive SAGB star. Both PTF11kx and SN~2002ic showed pre-interaction spectra which resembled a bright Type Ia SN, but only the former also showed the classical double peak in the NIR light curve like normal type Ia, casting some doubts to the nature of the latter.


In contrast, the values retrieved for SN~1997cy suggests an ejecta mass too massive for a thermonuclear explosion. The \ca\/ pseudo-bolometric light-curve is roughly a factor two dimmer than the bolometric. We can assume that all pseudo-bolometric light curves similar to that of \ca\/  in luminosity and time coverage are also a factor two dimmer than the bolometric and hence the kinetic energy retrieved by the model has to be multiplied by at least a factor two. This would suggest a core-collapse explosion for at least SN~1997cy and SN~1999E.  The high values of $\chi^2$ for these two objects are due to the absence of constraints for the first $\sim$70 days of evolution, also leading to lower values of masses and a ratio M$_{\rm ej}$/M$_{\rm Ni}$ too high for stripped envelope SNe, too. Indeed, this is a consequence of the limits of a simple code such as the aforementioned one.

The best fit for SN~2005gj is clearly unphysical since it requires a higher \ni\/ mass than the ejecta mass. However, since the late time temporal coverage is lower than \ca\/, the degeneracy is higher than the other cases. For example, an alternative model (blue dashed line in Figure~\ref{fig:allfit}) with M$_{\rm CSM}=3.8$ \M\/ and M$_{\rm Ni}=0.0$ \M\/ well describe the light-curve behaviour. This CSM mass is inconsistent with that stripped by the WD from its companion but still in the range of an AGB stars.

\begin{figure*}
\includegraphics[width=18cm]{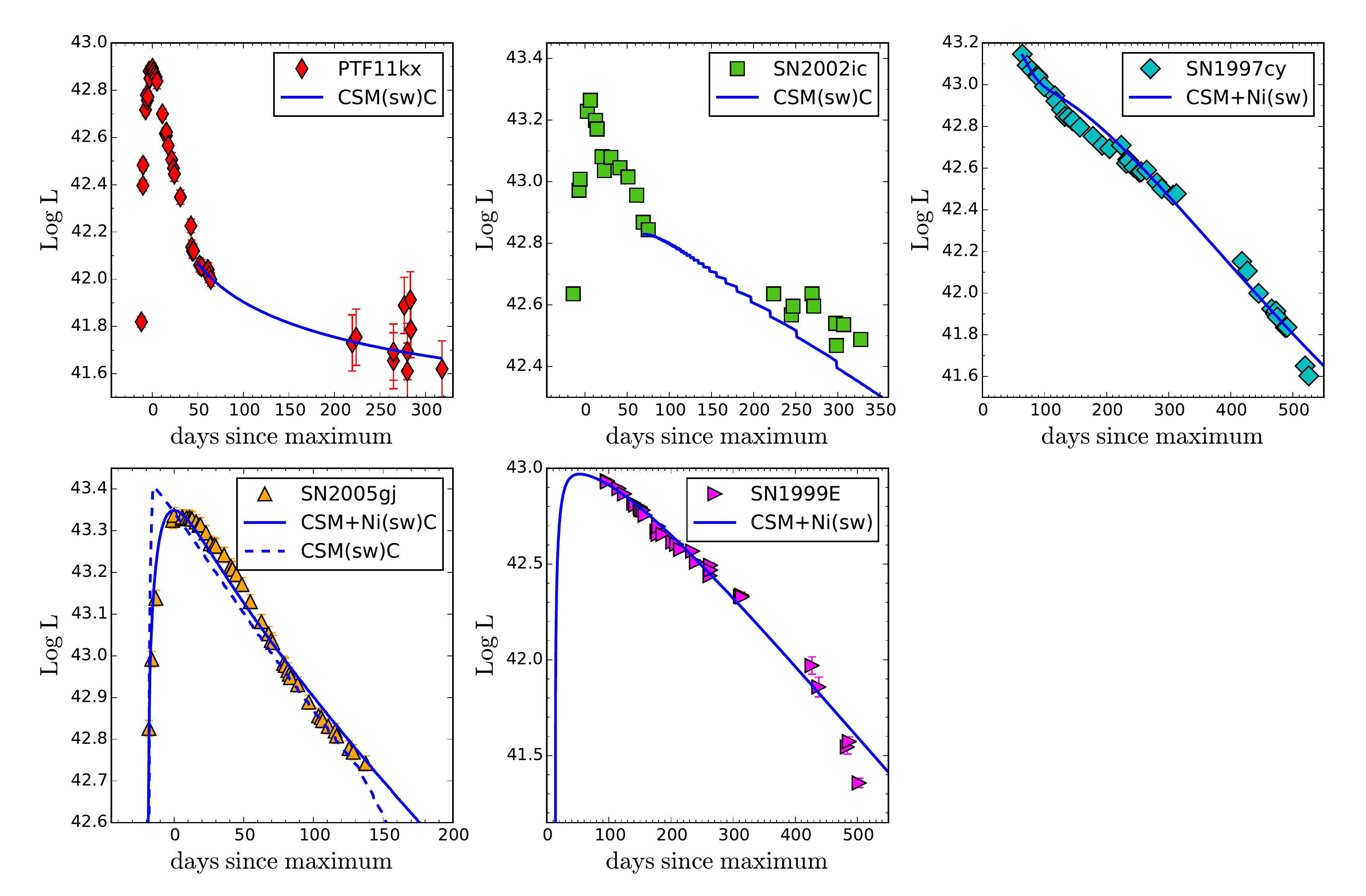}
\caption{Pseudo-bolometric light-curves of PTF11kx, SN~2002ic, SN~1997cy, SN~2005gj and SN~1999E and the semi-analytical model that best fit the light curve (blue solid line). The model used is reported in the legend. The suffix C refers to the model with constraints (see text).} 
\label{fig:allfit}
\end{figure*}

From the above analysis the only possible scenarios that can explain \ca\/ and similar type IIn/Ia-CSM objects are either a core collapse, or a thermonuclear explosion from a binary system which has a relatively massive CSM (such as from a WD + AGB star system). The latter has been already suggested by \citet{ha03} to explain the mass of the CSM in SN~2002ic.  A wind Roche-lobe overflow (RLOF) of $\sim50$ \kms\/  \citep[][]{pod07,moh14} - that happens when the companion fills its Roche lobe - can explain the process needed to account for a higher mass loss than what happens in normal WD + RG systems. A similar velocity was only observed in the unshocked CSM of PTF11kx \citep{di12}, while for \ca\/ and SN~1999E the velocities were a factor 3 or 4 higher. Furthermore, the RLOF can push the mass-loss up to a solar mass and hence lower than what was retrieved. An alternative suggestion could be that of a merger of a WD companion with the hot core of a massive AGB star leading to a SN Ia explosion that occurs at the end of the common envelope phase or shortly after \citep{so13}. However, this scenario would not explain the relatively high velocity of the unshocked wind observed in SNe 2012ca and 1999E.

\section{Conclusion}\label{sec:end}
We have presented extensive photometric and spectroscopic observations of the lowest redshift type IIn/Ia-CSM SN 2012ca. The light-curve for SN 2012ca covers the period of almost 600 days, likely including the peak epoch and comprises the first set of UV data obtained for a transient exhibiting such spectrophotometric behaviour.

The light-curve evolution of \ca\/ is similar to that of SNe~1997cy and 1999E after 100 days from maximum light, while the peak luminosity is fainter than that of SN~2002ic but brighter than PTF11kx. The spectroscopic coverage from prior to what has been identified as possible maximum light shows a persistent interaction with a CSM that lasts until 500 days. From the highest resolution spectra we also measured the velocity of the narrow absorption component associated with \Ha, and find it to be $v\sim200$ \kms\/. This is comparable to what found for SN~1999E by \citet{ri03} and a factor 3-4 higher than what was found for PTF11kx \citep[$v\sim65$ \kms][]{di12}.

From our spectroscopic analysis it is apparent that the spectra of \ca\/ are dominated exclusively by H, He, Fe and Ca, with some small contributions possible from O and Mg. That is also true for other type IIn/Ia-CSM. Moreover, the luminosity of \ca\/ at $\sim$1 year from peak, as well as those of SNe 1997cy, 1999E, 2002ic and 2005gj is more than 100 times brighter than that observed for a bright type Ia SN. Hence, only the CSM is responsible for the line profiles observed in \ca\/ from 1 year onward, as an underlying Type Ia SN would be simply too faint to be seen or contribute. We note that even if the peak epoch could not be well constrained, the peak would only have reached before our best estimate and that would strengthen an analysis based on the light curve.
Since we have shown that there is no spectral evolution in \ca\/ from the first epoch until the last, it appears reasonable that spectroscopic properties of \ca\/, as well as SNe 1997cy and 1999E are determined solely by the physical conditions within the CSM and the ejecta.

We also find that the UV contribution to the total bolometric flux is low ($<5$\%), which may indicate a flat ejecta density profile more similar to that of a H-free core-collapse SN. Furthermore, the total luminosity obtained by integrating over the entire light-curve of SN 2012ca implies a conversion efficiency of kinetic energy to luminosity of between 20 and 70  per cent in the case of a thermonuclear explosion (500\% to 2000\% in the case of the kinetic energy retrieved by the model), and 3 to 45 per cent in the case of a core-collapse (10\% to 20\% in the case of the kinetic energy retrieved by the model for a broad line type Ic), with the broad line CCSNe requiring the lowest conversion efficiency. We note that if the maximum of the lightcurve was even earlier than the date we adopted, these conversion efficiencies would need to be even higher.

We applied a semi-analytical code in order to estimate the ejecta and CSM masses, and the kinetic energy required to power the light-curves of \ca\/ and similar SNe. 
For \ca\/ we required even higher kinetic energies which would rule out the thermonuclear scenario, together with $0.9\lesssim M_{\rm ej}$(\M)$\lesssim 1.2$ and $2.4\lesssim M_{\rm CSM}$(\M)$\lesssim 2.6$. We retrieved similar values for the other five type IIn/Ia-CSM SNe with derived masses $0.4\lesssim M_{\rm ej}$(\M)$\lesssim 2.3$ and $0.4\lesssim M_{\rm CSM}$(\M)$\lesssim 4.4$. PTF11kx was the only object for which the modelling returned an ejecta mass and other parameters which were consistent with a thermonuclear explosion of a WD.

In the absence of an early period where the SN spectrum is not dominated by interaction (as for PTF11kx or SN~2002ic), it is difficult to unambiguously determine the true nature of a type IIn/Ia-CSM event. Even in the cases where such spectra {\it are} available, care must be taken in classifying the spectrum, as shown in the case of SN~2004aw, a type Ic SN which resembled a type Ia and hence was initially misclassified as the latter \citep[][]{ta06}.

Overall, the energy budget required for \ca\/ and the necessary high conversion efficiency of the SN kinetic energy to luminosity provides the strongest evidence for it being a core-collapse SN.
Secondary evidence for a core-collapse comes from the velocity of the narrow absorption component of \Ha\/, and the ejecta and CSM masses retrieved from semi-analytic modelling.

As shown in this work, energetic arguments can provide information on the nature of some of these events, given simple constraints such as that the luminosity from interaction must be less than the kinetic energy of the underlying SN. 
The existence of ostensibly similar (at least spectroscopically) objects such as PTF11kx, which has been unambiguously shown to be a type Ia SN, and SN 2012ca, for which a core-collapse is favoured based on the mass and energy budget, may point towards the existence of two physical channels for these transients which lead to observationally similar SNe. Such a result is also hinted at by the very different lightcurves exemplified by PTF11kx and \ca.

\section*{Acknowledgments}
CI thanks Stuart Sim, Bruno Leinbundgut, Maria Letizia Pumo and Andrea Pastorello for helpful discussions. CI and MF thanks Anders Jerkstrand for helpful comments. CI also thanks Melissa Graham for the LCOGT data acquisition.
This work is based on observations collected at the European Organisation for Astronomical Research in the Southern Hemisphere, Chile as part of PESSTO, (the Public ESO Spectroscopic Survey for Transient Objects Survey) ESO program 188.D-3003, 191.D-0935.
It is also based on observations taken at the Panchromatic Robotic Optical Monitoring and Polarimetry
Telescope (PROMPT) through the CNTAC proposal CN2012A-103; 
the Australian National
University 2.3m Telescope and the {\it Swift} satellite.
This work makes use of observations from the LCOGT network.
Funded by the European Research
Council under the European Union's Seventh Framework Programme
(FP7/2007-2013)/ERC Grant agreement n$^{\rm o}$ [291222] (SJS).
This work was partly supported by the European Union FP7 programme through ERC grant number 320360.
SB and AP  acknowledge the PRIN-INAF 2011 project ``Transient Universe: from ESO Large to PESSTO". 
Support for G.P. is provided by the Ministry of Economy, Development, and Tourism's Millennium Science Initiative through grant IC120009, awarded to The Millennium Institute of Astrophysics, MAS.
This research has made use of the NASA/IPAC Extragalactic
Database (NED) which is operated by the Jet Propulsion Laboratory,
California Institute of Technology, under contract with the National
Aeronautics and Space Administration.

\appendix

\section{Tables}

\begin{table*}
\caption{{\it UBVRI} magnitudes of \ca\/ with associated uncertainties in parentheses}
\begin{center}
\begin{tabular}{ccccccccc}
\hline
\hline
Date & MJD & Phase* & {\it U} & {\it B} & {\it V} & {\it R} & {\it I} & Inst.\\
yy/mm/dd &  & (days)& & & & & &\\
\hline
12/04/25  & 56042.58  &-5.6&  & & & 14.94 (0.03) &	&      BOSS\\
12/04/28 &  56045.58  & -2.6& & & & 14.78 (0.02) &	&      BOSS\\
12/04/30 & 56047.88  &  -0.3 &16.31 (0.08) & 16.15 (0.07) & 15.46 (0.05) & &	&SWIFT\\
12/05/02 &  56049.57 &  1.4 & & & & 14.77 (0.01) &	&      BOSS\\
12/05/03 & 56050.06 & 1.9 &	16.31 (0.13)	&	& & &&	SWIFT\\
12/05/04 & 56051.93 &3.7 &	16.32 (0.09) &	16.15 (0.07)	& 15.42 (0.06) & & &	SWIFT\\
12/05/05  & 56052.22 & 4.0&  & 16.15 (0.06) &  15.43 (0.03) & 14.82 (0.06) & 14.64 (0.08)   &PROMPT\\
12/05/06 & 56053.60 &5.4&	16.33 (0.09)	&16.16 (0.07)	& 15.43 (0.06) & & &	SWIFT\\
12/05/08 & 56055.12& 6.9&	16.35 (0.08)	&16.17 (0.07)	& 15.44 (0.06) & & &SWIFT\\
12/05/08  & 56055.22 &  7.0 & &16.17 (0.04) &15.46 (0.04) & 14.84 (0.05) & 14.68 (0.08)  &PROMPT\\
12/05/10 & 56057.48& 9.3 &	16.36 (0.08)	& 16.16 (0.07) & 15.46 (0.06) & &	&SWIFT\\
12/05/10 & 56057.91& 9.7 & 	& & & 14.88 (0.03) &	&      BOSS\\
12/05/12 & 56059.49& 11.3 &	16.36 (0.08)	&16.17 (0.07)	& 	15.47 (0.06) &&&	SWIFT\\
12/05/18 & 56065.29& 17.1 &	16.37 (0.08) &16.18 (0.07)	& 	15.48 (0.07) &&&	SWIFT\\
12/05/21 & 56068.03& 19.8 &	16.39 (0.08)	&16.18 (0.07)	& 	15.47 (0.06) &&&	SWIFT\\
12/05/24 & 56071.93& 23.7 &	16.40 (0.08)	&16.19 (0.07)	& 	15.48 (0.06) &&&	SWIFT\\
12/05/27 & 56074.49& 26.3 &	16.42 (0.10)	&16.19 (0.08)	& 	15.48 (0.08) 	&&&SWIFT\\
12/05/30 & 56077.52& 29.3 &	16.44 (0.08)	&16.20 (0.07) 	&15.49 (0.06)       &&&	SWIFT\\
12/06/11 & 56089.49& 41.3 &	 & & &15.02 (0.04) &	&      BOSS\\
12/06/26 & 56104.45& 56.2 &	 & & & 15.08 (0.05) &	   &   BOSS\\
12/06/27  & 56105.24 &  57.0 &  & 16.33 (0.08) & 15.58 (0.09) & 15.02 (0.10)& 14.88 (0.13)     &PROMPT\\
12/07/17  & 56125.23  & 77.0 & & 16.46 (0.10) &  15.65 (0.15) &15.28 (0.17) & 14.98 (0.17)    &PROMPT\\
12/07/19  & 56127.41 & 79.2 &  & & &15.25 (0.05) &	  &    BOSS\\
12/07/21  & 56129.98  & 81.8 & & & 15.67 (0.15) &15.29 (0.12) &14.91 (0.13)	   &  PROMPT\\
12/07/30 &  56138.07  & 89.9 & & 16.55 (0.08) & 15.69 (0.09)& 15.32 (0.09) & 15.04 (0.09)	&      PROMPT\\
12/08/24  & 56164.06  & 115.8 & 16.82 (0.04) & & & 15.50 (0.02) &    &NTT\\
12/08/26 &  56166.05   & 117.8 & 16.84 (0.03) & & &  &	      &NTT\\
12/09/09  & 56180.11  & 131.9 & 16.90 (0.10) & & & 15.61 (0.05) &	      &NTT\\
12/09/15 &  56186.01  & 137.8 & 16.92 (0.04) & & & 15.66 (0.03) &	      &NTT\\
12/10/08 &  56209.05   & 160.8 & 17.01 (0.16) & & & 15.77 (0.16) &		&NTT\\
12/10/15 &  56216.03   & 167.8 &17.02 (0.06) & & & 15.80 (0.03) &		&NTT\\
12/10/22 &  56223.02   & 174.8 &17.05 (0.06) & & & 15.86 (0.05) &	&NTT\\
12/11/06 &  56238.04  & 189.8 & 17.13 (0.07) & & &&	&	NTT\\
12/11/09 &  56240.40  & 192.2 & & & & 15.93 (0.07) &     & BOSS\\
12/11/11 &  56244.02  & 195.8 & 17.21 (0.09) & & & 15.96 (0.05) & 	&NTT\\
12/11/20 &  56252.01 &  203.8 & 17.25 (0.08) & & &16.00 (0.05) &	&NTT\\
13/03/18 &  56370.37 & 322.2 &	&  &  17.04 (0.04)  & &  &NTT\\
13/04/18 &  56401.35 & 353.1 &	& &  17.30 (0.05)  & &  &NTT\\
13/04/23 &  56405.64  &  357.4 & & & &    17.34 (0.07) &	&      BOSS\\
13/05/07 &  56419.61  & 371.4 & & & &   17.44 (0.06) &	  &    BOSS\\
13/05/28  & 56440.77  & 392.6 &  &  17.80 (0.12)& 17.56 (0.10) &  17.63 (0.11) & 17.79 (0.12)  &  FTS\\
13/05/28 &  56441.12  & 392.9 &  &   &  17.55 (0.04) &  17.63 (0.03) & 17.76 (0.04)  &LCOGT\\
13/05/30 &  56443.12  &  394.9 & & &  17.63 (0.06)  & 17.66 (0.06) & & LCOGT\\
13/06/02 &  56446.30  & 398.1 &  & & 17.70 (0.05) & 17.74 (0.07)  & 17.86 (0.13) & LCOGT\\
13/06/03 &  56447.31  &  399.1 &  & &  17.72 (0.04) & 17.72 (0.04)  & 17.87 (0.04) &  LCOGT\\
13/06/04 &  56448.31  & 400.1 &  & &  17.74 (0.04) & 17.75 (0.10)   &  17.88 (0.12) & LCOGT\\
13/06/05 &  56449.30  &  401.1 & & &  17.75 (0.08) & 17.76 (0.12) &  & LCOGT\\
13/06/06 &  56450.31  &  402.1 & & &  17.76 (0.09) & 17.78 (0.15)  & 17.85 (0.13) & LCOGT\\
13/06/11 &  56455.31  & 407.1 &  & &   17.79 (0.10) & 17.85 (0.15) & 17.94 (0.15) & LCOGT\\
13/06/13 &  56457.31  & 409.1 &   & &  17.83 (0.09) &  18.03 (0.13)  & & LCOGT\\
13/07/18 &  56491.36  & 443.2 &  &   18.51 (0.17)  &  18.37 (0.16) & 18.43 (0.20) &  &FTS\\
13/07/24 &  56497.56 &  449.4 & &   18.57 (0.12) &  18.60 (0.05)  & 18.62 (0.06) &  &FTS\\
13/08/09 &  56514.18 & 466.0 & & & &  18.82 (0.20)  &&SMARTS\\
13/08/12 &  56517.18 & 469.0 & & & & &  18.86 (0.28) & SMARTS\\
13/08/16 &  56521.04 & 472.8 & & & 19.31 (0.06) & & & NTT\\
13/09/11 &  56547.11 & 498.9 & & &  19.79 (0.08)  & & &  NTT\\
13/10/07 & 56573.02 &  524.8 & & 21.75 (0.11) & 20.75 (0.12) & 20.79 (0.14) & &NTT\\
13/10/13 & 56579.01 &  530.8 & &  & 21.27 (0.24) & & &NTT\\
13/11/11 & 56608.05 & 559.8& & & $>$21.50 & $>$ 21.30 & & NTT\\
\hline
\end{tabular}
\end{center}
* Phase with respect to the {\it R}-band maximum.
\label{table:snm}
\end{table*}%

\begin{table*}
\caption{{\it griz} magnitudes of \ca\/ with associated uncertainties in parentheses.}
\begin{center}
\begin{tabular}{cccccccc}
\hline
\hline
Date & MJD & Phase* & {\it g} & {\it r} & {\it i} & {\it z} & Inst.\\
yy/mm/dd &  &(days)  & & & & &\\
\hline
12/04/28  & 56046.40  & -1.8 &  15.17 (0.02)  & & &  &NTT\\
12/05/05  & 56052.22 & 4.0&  14.83 (0.03) & 14.78  (0.07) & 14.40 (0.07) & 14.37 (0.09)	      &PROMPT\\
12/05/08  & 56055.22 &  7.0 &  14.88 (0.03) & 14.79 (0.08) & 14.46 (0.07) & 14.39 (0.09)	      &PROMPT\\
12/06/27  & 56105.24 &  57.0 & & 14.99 (0.09) & 14.50 (0.11) & 14.64 (0.13)	      &PROMPT\\
12/07/07  & 56115.11  & 66.9 &  & 15.10 (0.16) & 14.62 (0.14) &14.74 (0.15)	    &  PROMPT\\
12/07/17  & 56125.23  & 77.0 & 15.51 (0.20) & 15.13 (0.17) & 14.71 (0.20) &	      &PROMPT\\
12/07/24 &  56132.26  & 84.1 && 15.16 (0.06) &14.71 (0.07) &14.85 (0.07)	    &  PROMPT\\
12/07/31 &  56139.22  & 91.0 && 15.28 (0.08) & 14.84 (0.10) & 15.02 (0.06)	   &   PROMPT\\
12/08/08  & 56148.09  & 99.9 &  15.76 (0.03) & 15.31 (0.03) & 14.85 (0.03)& 15.05 (0.03) &	      NTT\\
12/08/18 &  56157.22  & 109.0 &  15.82 (0.07) & 15.35 (0.06) & 14.93 (0.10) & 15.10 (0.09)	 &     NTT\\
12/08/21  & 56160.17  & 112.0 &  15.84 (0.05) & 15.36 (0.05) & 14.93 (0.07) & 15.13 (0.10)	 &     PROMPT\\
12/08/24  & 56164.06  & 115.8 &  15.88 (0.04) & 15.36 (0.03) & 14.95 (0.03) & 15.14 (0.04)	      &NTT\\
12/08/26 &  56166.05   & 117.8 & 15.89 (0.04)& 15.34 (0.06)& 14.94 (0.04) & 15.15 (0.09)	      &NTT\\
12/09/09  & 56180.11  & 131.9 & 15.95 (0.04) &  15.43 (0.05) & 15.02 (0.05) &15.20 (0.04)	      &NTT\\
12/09/15 &  56186.01  & 137.8 & 15.99 (0.03) &  15.48 (0.04) & 15.12 (0.03)& 15.26 (0.03)	      &NTT\\
12/10/08 &  56209.05   & 160.8 & 16.05 (0.05) &15.57 (0.07) & 15.29 (0.04) &15.47 (0.08)		&NTT\\
12/10/15 &  56216.03   & 167.8 & 16.06 (0.04) &  15.63 (0.04) & 15.30 (0.02) &15.49 (0.03)		&NTT\\
12/10/22 &  56223.02   & 174.8 &16.12 (0.03) & 15.67 (0.04) & 15.40 (0.03) &15.51 (0.02)		&NTT\\
12/11/06 &  56238.04  & 189.8 &  16.16 (0.03) &  15.76 (0.05) &15.48 (0.02)& 15.64 (0.02)	&	NTT\\
12/11/11 &  56244.02  & 195.8 &  16.17 (0.03) & 15.80 (0.04) & 15.52 (0.02) &15.66 (0.02)		&NTT\\
12/11/20 &  56252.01 &  203.8 &  16.20 (0.04) & 15.82 (0.04) & 15.52 (0.02) &15.72 (0.03)		&NTT\\
13/03/10  & 56361.67 & 313.5 &  16.98 (0.10)  &  16.52 (0.09) &  16.11 (0.11) & 16.15 (0.08) & NTT\\
13/03/11 &  56362.67 & 314.5 & 16.96 (0.11)   & 16.54 (0.11)  &  16.18 (0.12)  &16.18 (0.09) & NTT\\
13/03/31 &  56382.63 & 334.4 & 17.39 (0.12)  &  16.81 (0.13)  &  16.90 (0.14)  & 16.50 (0.13)  &NTT\\
13/10/07 & 56573.02 &  524.8 & 20.90 (0.07) & 20.82 (0.10) &  20.54 (0.12) & 20.83 (0.10) & NTT\\
13/11/11 & 56608.05 & 559.8& & $>$21.30 & $>$21.10 & $>$21.20 & NTT\\
\hline
\end{tabular}
\end{center}
* Phase with respect to the {\it R}-band maximum.
\label{table:sns}
\end{table*}%

\begin{table*}
\caption{{\it Swift}+UVOT UV magnitudes of \ca\/ in the Vega system. Associated errors in parentheses.}
\begin{center}
\begin{tabular}{cccccc}
\hline
\hline
Date & MJD & Phase* & {\it uvw2} & {\it uvm2} & {\it uvw1} \\
yy/mm/dd &  &(days)  &  & &\\
\hline
12/04/30 &	56047.88 	 &-0.3&17.85 (0.12)&	17.49 (0.11)&	16.94 (0.10) \\
12/05/03 & 	56050.06 	 &1.9 &&	&	16.79 (0.10) \\
12/05/04 & 	56051.93 	 &3.7&17.99 (0.12)&	17.87 (0.11)&	16.83 (0.10) \\
12/05/06 & 	56053.60 	 &5.4&17.96 (0.12)&	17.62 (0.11)&	16.92 (0.10) \\
12/05/08 & 	56055.12 	 &6.9&17.99 (0.12)&	17.49 (0.10)&	16.94 (0.10) \\
12/05/10 & 	56057.48 	 &9.3&17.89 (0.12)&	17.61 (0.11)	&16.86 (0.10) \\
12/05/12 & 	56059.49  &11.3&	18.08 (0.12)&	17.80 (0.11)	&16.91( 0.10) \\
12/05/18 & 	56065.29  &17.1&	17.85 (0.12)&	17.78 (0.12)&	16.90 (0.09)\\
12/05/21 & 	56068.03 	 &19.8&17.92 (0.12)&	17.66 (0.11)&	16.97 (0.09)\\
12/05/24 & 	56071.93  &23.7&	17.82 (0.12) &	17.62 (0.10)&	16.95 (0.10) \\
12/05/27 & 	56074.49  &26.3&	17.86 (0.15)	&17.65 (0.13)&	17.06 (0.12)\\
12/05/30 & 	56077.52 & 29.3&	17.85 (0.12)	&17.70 (0.13)&	16.94 (0.09) \\
\hline
\end{tabular}
\end{center}
\label{table:snuv}
* Phase with respect to the {\it R}-band maximum.
\end{table*}%

\begin{table*}
\caption{NIR SOFI photometry of \ca\/ in the 2MASS system.  Errors are given in parentheses.}
\begin{center}
\begin{tabular}{cccccc}
\hline
\hline
Date & MJD & Phase* & {\it J} & {\it H} & {\it K} \\
yy/mm/dd &  & (days)  &  & &\\
\hline
12/08/09 &  56149.20 &101.0 &	 14.96  (0.04)  &    14.69  (0.04) &	   14.77  (0.04)\\
12/09/15  & 56186.03 & 137.8&	 15.11  (0.04)  &    15.01  (0.08) &	   14.89  (0.10)\\
12/11/14 &  56246.01 & 197.8&	 15.58  (0.04)   &   15.57  (0.04) &	   15.45  (0.04)\\
13/03/05  & 56357.37 & 309.2&	 16.37  (0.05)   &   16.14  (0.07) &	   15.61  (0.04)\\
13/04/11  & 56394.28 & 346.1&	 16.70  (0.08)   &   16.21  (0.11)	 &   15.43  (0.07)\\
13/10/14 & 56580.02& 531.8 & 19.25  (0.19) &     17.08  (0.09)	&   15.30  (0.07)\\
13/11/02 & 56598.99& 550.8 & 19.37  (0.34) &     17.26  (0.07)	&   15.38  (0.10)\\
\hline
\end{tabular}
\end{center}
* Phase with respect to the {\it R}-band maximum.
\label{table:snir}
\end{table*}%

\begin{table*}
\caption{{\it UBVRI} magnitudes of local sequence stars in the field of \ca.}
\begin{center}
\begin{tabular}{lccccc}
\hline
\hline
Star & {\it U} & {\it B} & {\it V} & {\it R} & {\it I} \\
\hline
1 & 16.10 (0.01) & 15.59 (0.01) & 15.50 (0.03) & 15.12 (0.02) & 15.54 (0.02)\\
2 & 18.01 (0.03) & 17.31 (0.02) & 17.07 (0.05) & 16.60 (0.02) & 17.01 (0.03)\\
3 & 18.30 (0.01) & 17.70 (0.01) & 17.48 (0.02) & 17.10 (0.01) & 17.13 (0.02)\\
4 & 16.63 (0.01) & 16.33 (0.01) & 16.20 (0.02) & 15.89 (0.01) & 15.92 (0.03)\\
5 & 19.31 (0.03) & 18.32 (0.02) & 17.87 (0.01) & 17.31 (0.02) & 17.19 (0.02)\\
6 & 15.72 (0.01) & 14.84 (0.01) & 14.45 (0.03) & 14.02 (0.01) & 14.00 (0.01)\\
7 & 17.46 (0.02) & 16.99 (0.02) & 16.74 (0.03) & 16.37 (0.01) & 16.36 (0.02)\\
8 & 19.00 (0.02) & 18.34 (0.04) & 18.05 (0.01) & 17.65 (0.02) & 17.60 (0.02)\\
9 & 19.04 (0.04) & 18.51 (0.03) & 18.20 (0.02) &17.82 (0.01) & 17.78 (0.02)\\
10 & 20.42 (0.06) & 18.97 (0.05) & 18.36 (0.06) & 17.76 (0.02) & 17.60 (0.02)\\
11 & 19.95 (0.03) &18.35 (0.04) & 17.33 (0.04) & 16.36 (0.01) & 15.58 (0.01)\\
12 & 19.70 (0.03) & 18.27 (0.04) & 17.63 (0.04) & 16.96 (0.01) & 16.72 (0.02)\\
\hline
\end{tabular}
\end{center}
\label{table:ssj}
\end{table*}%

\begin{table*}
\caption{{\it griz} magnitudes of local sequence stars in the field of \ca.}
\begin{center}
\begin{tabular}{lcccc}
\hline
\hline
Star & {\it g} & {\it r} & {\it i} & {\it z} \\
\hline
1 & 15.60 (0.01) & 15.17 (0.01) & 14.50 (0.01) & 14.90 (0.01)\\
2 & 17.20 (0.01) & 16.63 (0.01) & 15.91 (0.01) & 16.25 (0.02)\\
3 & 17.64 (0.01) & 17.15 (0.01) & 16.46 (0.01) & 16.84 (0.01)\\
4 & 16.33 (0.01) & 15.92 (0.01) & 15.27 (0.01) & 15.67 (0.01)\\
5 & 18.10 (0.01) & 17.28 (0.01) & 16.51 (0.01) & 16.81 (0.01)\\
6 & 14.65 (0.01) & 14.05 (0.01) & 13.32 (0.01) & 13.66 (0.01)\\
7 & 16.89 (0.01) & 16.37 (0.01) & 15.66 (0.01) & 15.99 (0.01)\\
8 & 18.22 (0.01) & 17.69 (0.01) & 16.96 (0.01) & 17.33 (0.01)\\
9 & 18.38 (0.01) & 17.78 (0.01) & 17.10 (0.01) &17.39 (0.02)\\
10 & 18.65 (0.01) & 17.74 (0.01) & 16.91 (0.01) & 17.17 (0.01)\\
11 & 17.80 (0.01) &16.36 (0.01) & 14.88 (0.01) & 14.99 (0.01)\\
12 & 17.95 (0.01) & 16.93 (0.01) & 16.02 (0.01) & 16.26 (0.01)\\
\hline
\end{tabular}
\end{center}
\label{table:sss}
\end{table*}%

\begin{table*}
\caption{Journal of spectroscopic observations.}
\begin{center}
\begin{tabular}{cccccl}
\hline
\hline
Date & MJD & Phase* &Range & Resolution & Instrumental  \\
yy/mm/dd &  & (days)  &(\AA)  & (\AA) &Configuration\\
\hline
12/04/28 &       56046.40 & -1.8 &3585--9070 & 30 & NTT+EFOSC2+gm13\\                  
12/05/03&       56051.41        & 3.2   & 4550--8730  & 3,4 & Gemini+GMOS+B600/R400         \\                                                                              
12/05/31&       56079.40          &  31.2  &  4550--8730  & 3,4  & Gemini+GMOS+B600/R400    \\                                                                                
12/07/07&        56116.42          &  68.2  & 4550--8230   & 3,4 & Gemini+GMOS+B600/R400 \\                                                                                
12/08/08&        56148.06 & 99.8 &3585--9070 & 18 & NTT+EFOSC2+gm13\\                       
12/08/09   &       56149.23 &101.0 &9180--13220&   23 & NTT+SOFI+BG\\                                                       
12/08/10 &         56150.30  &102.1&3430--9370&2 & ANU 2.3m+WiFeS+B300/R300\\                                                                                        
12/08/17 &       56157.18&  109.0  & 3585--9070  & 18 & NTT+EFOSC2+gm13\\                                      
12/08/24 &       56164.02& 115.8 &3585--9070   & 18 & NTT+EFOSC2+gm13\\                      
12/09/06  &            56172.30 & 124.1 &3430--9370&2 & ANU 2.3m+WiFeS+B300/R300\\                                                                                               
12/09/07  &     56173.30        & 125.1 &3430--9370&2 & ANU 2.3m+WiFeS+B300/R300\\                                                                                                        
12/09/15   &        56185.98 & 137.8     &  3585--9070 & 18 & NTT+EFOSC2+gm13\\                   
12/09/15 &       56186.05 &  137.8  &  9180--13220 &   23 & NTT+SOFI+BG\\                                                    
12/09/22   &      56193.11       &  144.9   &3430--9370&2 & ANU 2.3m+WiFeS+B300/R300\\                                     
12/09/23   &       56194.10      &  145.9   &3430--9370&2 & ANU 2.3m+WiFeS+B300/R300\\                                                                                            
12/09/24  &       56195.01&  146.8  &14600--24710 &   33 & NTT+SOFI+RG\\                                                    
12/10/22  &      56223.06 & 174.9 & 9180--13220&   23 & NTT+SOFI+BG\\                                                                
12/11/12  &      56244.04 &195.8 &3585--9070 & 18 & NTT+EFOSC2+gm13\\                             
12/11/13   &       56245.00 &196.8 & 9180--13220   &   23 & NTT+SOFI+BG\\                                 
13/02/20  &      56344.34 & 296.1 &3585--9070 & 18 & NTT+EFOSC2+gm13\\                             
13/03/11   &      56363.28 & 315.1 &  9180--24710 &   23,33 & NTT+SOFI+BG/RG\\                                                          
13/03/18 &       56370.38 &322.2&3585--9070 & 18 & NTT+EFOSC2+gm13\\                             
13/04/18&        56401.35& 353.1 &3585--9070 & 18 & NTT+EFOSC2+gm13\\                             
13/05/08&        56421.20              &  373.0 &5280--7060&2 & ANU 2.3m+WiFeS+R300\\                                                                                   
13/05/26&        56439.20         & 391.0  &3430--9370&2 & ANU 2.3m+WiFeS+B300/R300\\                                                                                                 
13/06/03&        56446.20        & 398.0    &3430--6890&2 & ANU 2.3m+WiFeS+B300/R700\\                                                                                        
13/08/16&        56521.04& 472.8 &3585--9070 & 18 & NTT+EFOSC2+gm13\\                             
13/09/11     &      56547.11& 498.9 &3585--9070 & 18 & NTT+EFOSC2+gm13\\       
13/10/13     &      56579.05&530.8 &3585--9070 & 18 & NTT+EFOSC2+gm13\\                              
\hline
\end{tabular}
\end{center}
* Phase with respect to the {\it R}-band maximum.
\label{table:sp}
\end{table*}%


\begin{thebibliography}{99}\label{bib}
\bibitem[\protect\citeauthoryear{Aldering et 
al.}{2006}]{al06} Aldering G., et al., 2006, ApJ, 650, 510
\bibitem[\protect\citeauthoryear{Arnett 
\& Fu}{1989}]{ar89} Arnett W.~D., Fu A., 1989, ApJ, 340, 396
\bibitem[\protect\citeauthoryear{Benetti et 
al.}{2006}]{be06} Benetti S., Cappellaro E., Turatto M., 
Taubenberger S., Harutyunyan A., Valenti S., 2006, ApJ, 653, L129 
\bibitem[\protect\citeauthoryear{Brown}{2013}]{br13} Brown 
T., 2013, POBeo, 92, 91 
\bibitem[\protect\citeauthoryear{Chandra et 
al.}{2015}]{cha15} Chandra P., Chevalier R.~A., Chugai N., 
Fransson C., Soderberg A.~M., 2015, ApJ, 810, 32
\bibitem[\protect\citeauthoryear{Chatzopoulos, Wheeler, 
\& Vinko}{2012}]{cha12} Chatzopoulos E., Wheeler J.~C., Vinko J., 2012, ApJ, 746, 121
\bibitem[\protect\citeauthoryear{Chevalier 
\& Fransson}{1994}]{cf94} Chevalier R.~A., Fransson C., 1994, ApJ, 420, 268
\bibitem[\protect\citeauthoryear{Chevalier 
\& Irwin}{2011}]{ch11} Chevalier R.~A., Irwin C.~M., 2011, ApJ, 729, L6
\bibitem[\protect\citeauthoryear{Childress et 
al.}{2015}]{chi15} Childress M.~J., et al., 2015, arXiv, 
arXiv:1507.02501
\bibitem[\protect\citeauthoryear{Childress et al.}{2014}]{Chi14}
Childress, M.~J., Vogt, F.~P.~A., Nielsen, J., \& Sharp, R.~G.\ 2014, Ap\&SS,, 349, 617 
\bibitem[\protect\citeauthoryear{Deng et al.}{2004}]{de04} 
Deng J., et al., 2004, ApJ, 605, L37 
\bibitem[\protect\citeauthoryear{Dilday et al.}{2012}]{di12} 
Dilday B., et al., 2012, Sci, 337, 942
\bibitem[\protect\citeauthoryear{Dopita et 
al.}{2010}]{do10} Dopita M., et al., 2010, Ap\&SS, 327, 245
\bibitem[\protect\citeauthoryear{Drescher, Parker, 
\& Brimacombe}{2012}]{dr12} Drescher C., Parker S., Brimacombe J., 2012,  CBET,3101,1
\bibitem[\protect\citeauthoryear{Dwarkadas \& Gruszko}{2012}]{dw12}
Dwarkadas, V.~V., Gruszko, J.\ 2012, MNRAS, 419, 1515 
\bibitem[\protect\citeauthoryear{Filippenko}{1997}]{fi97} Filippenko A.~V., 1997, ARA\&A, 35, 309 
\bibitem[\protect\citeauthoryear{Foreman-Mackey et 
al.}{2013}]{For13} Foreman-Mackey D., Hogg D.~W., Lang D., 
Goodman J., 2013, PASP, 125, 306
\bibitem[\protect\citeauthoryear{Fox et al.}{2015}]{fox15} 
Fox O.~D., et al., 2015, MNRAS, 447, 772 
\bibitem[\protect\citeauthoryear{Fox 
\& Filippenko}{2013}]{fox13} Fox O.~D., Filippenko A.~V., 2013, ApJ, 772, L6
\bibitem[\protect\citeauthoryear{Fransson et al.}{2014}]{fran14} 
Fransson C., et al.\ 2014, ApJ, 797, 118
\bibitem[\protect\citeauthoryear{Fraser et al.}{2013}]{fr13} 
Fraser M. et al.\ 2013, MNRAS, 433, 1312 
\bibitem[\protect\citeauthoryear{Fraser et al.}{2015}]{fr15} 
Fraser M., et al., 2015, MNRAS, 453, 3886 
\bibitem[\protect\citeauthoryear{Germany et 
al.}{2000}]{ge00} Germany L.~M., Reiss D.~J., Sadler E.~M., 
Schmidt B.~P., Stubbs C.~W., 2000, ApJ, 533, 320
\bibitem[\protect\citeauthoryear{Hamuy et al.}{2003}]{ha03} 
Hamuy M., et al., 2003, Natur, 424, 651
\bibitem[\protect\citeauthoryear{Hillebrandt \& Niemeyer}{2000}]{hn00} Hillebrandt W., Niemeyer J.~C., 2000, ARA\&A, 38, 191
\bibitem[\protect\citeauthoryear{Hunter et 
al.}{2009}]{hu09} Hunter D.~J., et al., 2009, A\&A, 508, 371
\bibitem[\protect\citeauthoryear{Inserra et 
al.}{2015}]{in15} Inserra C., et al., 2016, ApJ, in prep.
\bibitem[\protect\citeauthoryear{Inserra et 
al.}{2014}]{in14} Inserra C., et al., 2014, MNRAS, 437, L51
\bibitem[\protect\citeauthoryear{Inserra et 
al.}{2012}]{in12} Inserra C., et al., 2012, CBET, 3101, 1
\bibitem[\protect\citeauthoryear{Iwamoto et 
al.}{1998}]{iw98} Iwamoto K., et al., 1998, Natur, 395, 672 
\bibitem[Janka(2012)]{2012ARNPS..62..407J} Janka, H.-T.\ 2012, Annu. Rev. Nucl. Part. Sci., 62, 407 
\bibitem[\protect\citeauthoryear{Jerkstrand et 
al.}{2012}]{je12} Jerkstrand A., Fransson C., Maguire K., Smartt S., Ergon M., Spyromilio J., 2012, A\&A, 546, A28
\bibitem[\protect\citeauthoryear{Kennicutt, Tamblyn, 
\& Congdon}{1994}]{ken94} Kennicutt R.~C., Jr., Tamblyn P., Congdon C.~E., 1994, ApJ, 435, 22 
\bibitem[\protect\citeauthoryear{Kotak et al.}{2004}]{ko04} 
Kotak R., Meikle W.~P.~S., Adamson A., Leggett S.~K., 2004, MNRAS, 354, L13 
\bibitem[\protect\citeauthoryear{Landolt}{1992}]{la92} 
Landolt A.~U., 1992, AJ, 104, 340 
\bibitem[\protect\citeauthoryear{Leloudas et al.}{2015}]{le15}
Leloudas G., et al., 2015, A\&A, 574, A61 
\bibitem[\protect\citeauthoryear{Lira et al.}{1998}]{li98} 
Lira P., et al., 1998, AJ, 115, 234
\bibitem[\protect\citeauthoryear{Liu et 
al.}{2012}]{liu12} Liu Z.~W., Pakmor R., R{\"o}pke F.~K., Edelmann P., Wang B., Kromer M., Hillebrandt W., Han Z.~W., 2012, A\&A, 548, A2
\bibitem[\protect\citeauthoryear{Madau, Pozzetti, 
\& Dickinson}{1998}]{mad98} Madau P., Pozzetti L., Dickinson M., 1998, ApJ, 498, 106
\bibitem[\protect\citeauthoryear{Mazzali et 
al.}{2007}]{ma07} Mazzali P.~A., R{\"o}pke F.~K., Benetti 
S., Hillebrandt W., 2007, Sci, 315, 825
\bibitem[\protect\citeauthoryear{Matzner 
\& McKee}{1999}]{mat99} Matzner C.~D., McKee C.~F., 1999, ApJ, 510, 379
\bibitem[\protect\citeauthoryear{Meynet et 
al.}{2015}]{mey15} Meynet G., et al., 2015, A\&A, 575, A60 
\bibitem[\protect\citeauthoryear{Mohamed et 
al.}{2014}]{moh14} Mohamed S., et al., 2014, apn6.conf, 60
\bibitem[\protect\citeauthoryear{Ofek et al.}{2014}]{of14} 
Ofek E.~O., et al., 2014, ApJ, 781, 42
\bibitem[\protect\citeauthoryear{Nomoto, Thielemann, 
\& Yokoi}{1984}]{no84} Nomoto K., Thielemann F.-K., Yokoi K., 1984, ApJ, 286, 644
\bibitem[\protect\citeauthoryear{Nicholl et 
al.}{2014}]{ni14} Nicholl M., et al., 2014, MNRAS, 444, 2096
\bibitem[\protect\citeauthoryear{Pan, Ricker, 
\& Taam}{2010}]{pan10} Pan K.-C., Ricker P.~M., Taam R.~E., 2010, ApJ, 715, 78 
\bibitem[\protect\citeauthoryear{Pan, Ricker, 
\& Taam}{2012}]{pan12} Pan K.-C., Ricker P.~M., Taam R.~E., 2012, ApJ, 750, 151 
\bibitem[\protect\citeauthoryear{Patat et 
al.}{2011}]{pa11} Patat F., Chugai N.~N., Podsiadlowski P., Mason E., Melo C., Pasquini L., 2011, A\&A, 530, A63 
\bibitem[\protect\citeauthoryear{Podsiadlowski 
\& Mohamed}{2007}]{pod07} Podsiadlowski P., Mohamed S., 2007, BaltA, 16, 26
\bibitem[\protect\citeauthoryear{Poole et al.}{2008}]{po08} 
Poole T.~S., et al., 2008, MNRAS, 383, 627
\bibitem[\protect\citeauthoryear{Pignata et 
al.}{2004}]{pi04} Pignata G., et al., 2004, MNRAS, 355, 178
\bibitem[\protect\citeauthoryear{Prieto et al.}{2007}]{pr07} 
Prieto J.~L., et al., 2007, arXiv, arXiv:0706.4088
\bibitem[\protect\citeauthoryear{Pumo et al.}{2009}]{pu09} 
Pumo M.~L., et al., 2009, ApJ, 705, L138
\bibitem[\protect\citeauthoryear{Reichart et 
al.}{2005}]{re05} Reichart D., et al., 2005, NCimC, 28, 767
\bibitem[\protect\citeauthoryear{Rigon et al.}{2003}]{ri03} 
Rigon L., et al., 2003, MNRAS, 340, 191 
\bibitem[\protect\citeauthoryear{R{\"o}pke et 
al.}{2007}]{ro07} R{\"o}pke F.~K., Hillebrandt W., Schmidt 
W., Niemeyer J.~C., Blinnikov S.~I., Mazzali P.~A., 2007, ApJ, 668, 1132
\bibitem[\protect\citeauthoryear{Scalzo et al.}{2014}]{sc14} 
Scalzo R., et al., 2014, MNRAS, 440, 1498
\bibitem[\protect\citeauthoryear{Schlafly 
\& Finkbeiner}{2011}]{sf11} Schlafly E.~F., Finkbeiner D.~P., 2011, ApJ, 737, 103
\bibitem[\protect\citeauthoryear{Schlegel}{1990}]{sc90} Schlegel E.~M., 1990, MNRAS, 244, 269 
\bibitem[\protect\citeauthoryear{Silverman et 
al.}{2013a}]{si13a} Silverman J.~M., et al., 2013a, ApJS, 207, 
3
\bibitem[\protect\citeauthoryear{Silverman et 
al.}{2013b}]{si13b} Silverman J.~M., et al., 2013b, ApJ, 772, 
125 
\bibitem[\protect\citeauthoryear{Sim et al.}{2012}]{sim12} 
Sim S.~A., Fink M., Kromer M., R{\"o}pke F.~K., Ruiter A.~J., Hillebrandt 
W., 2012, MNRAS, 420, 3003
\bibitem[\protect\citeauthoryear{Skrutskie et 
al.}{2006}]{sk06} Skrutskie M.~F., et al., 2006, AJ, 131, 
1163
\bibitem[\protect\citeauthoryear{Smartt et 
al.}{2015}]{sm14} Smartt S.~J., et al., 2015, A\&A, 579, A40
\bibitem[\protect\citeauthoryear{Smith et al.}{2009}]{sm09} 
Smith N., et al., 2009, ApJ, 695, 1334
\bibitem[\protect\citeauthoryear{Soker et al.}{2013}]{so13} 
Soker N., Kashi A., Garc{\'{\i}}a-Berro E., Torres S., Camacho J., 2013, 
MNRAS, 431, 1541
\bibitem[\protect\citeauthoryear{Sternberg et 
al.}{2011}]{st11} Sternberg A., et al., 2011, Sci, 333, 856
\bibitem[\protect\citeauthoryear{Stritzinger et 
al.}{2002}]{st02} Stritzinger M., et al., 2002, AJ, 124, 
2100
\bibitem[\protect\citeauthoryear{Svirski, Nakar, 
\& Sari}{2012}]{svi12} Svirski G., Nakar E., Sari R., 2012, ApJ, 759, 108
\bibitem[\protect\citeauthoryear{Taddia et 
al.}{2012}]{ta12} Taddia F., et al., 2012, A\&A, 545, L7
\bibitem[\protect\citeauthoryear{Taubenberger et 
al.}{2006}]{ta06} Taubenberger S., et al., 2006, MNRAS, 371, 
1459
\bibitem[\protect\citeauthoryear{Taubenberger et 
al.}{2011}]{tau11} Taubenberger S., et al., 2011, MNRAS, 412, 
2735 
\bibitem[\protect\citeauthoryear{Trundle et 
al.}{2008}]{tr08} Trundle C., Kotak R., Vink J.~S., Meikle W.~P.~S., 2008, A\&A, 483, L47
\bibitem[\protect\citeauthoryear{Turatto et 
al.}{2000}]{tu00} Turatto M., et al., 2000, ApJ, 534, L57
\bibitem[\protect\citeauthoryear{Umeda 
\& Nomoto}{2008}]{un08} Umeda H., Nomoto K., 2008, ApJ, 673, 1014 
\bibitem[\protect\citeauthoryear{Valenti et 
al.}{2012}]{va12} Valenti S., et al., 2012, ATel, 4076, 1
\bibitem[\protect\citeauthoryear{Wang et al.}{2004}]{wa02} 
Wang L., Baade D., H{\"o}flich P., Wheeler J.~C., Kawabata K., Nomoto K., 
2004, ApJ, 604, L53
\bibitem[\protect\citeauthoryear{Wood-Vasey, Wang, 
\& Aldering}{2004}]{wv04} Wood-Vasey W.~M., Wang L., Aldering G., 2004, ApJ, 616, 339
\bibitem[\protect\citeauthoryear{Yaron 
\& Gal-Yam}{2012}]{ya12} Yaron O., Gal-Yam A., 2012, PASP, 124, 668 
\bibitem[\protect\citeauthoryear{Zhang et al.}{2012}]{zh12} 
Zhang T., et al., 2012, AJ, 144, 131 
\end{thebibliography}
\end{document}